\documentclass[fleqn,usenatbib]{mnras}

\DeclareRobustCommand{\VAN}[3]{#2}
\let\VANthebibliography\thebibliography
\def\thebibliography{\DeclareRobustCommand{\VAN}[3]{##3}\VANthebibliography}

\usepackage{graphicx}	
\usepackage{amsmath}	
\usepackage{amssymb}	
\usepackage{xcolor}
\usepackage{graphics, setspace}
\usepackage{cancel}

\newcommand{\mathsym}[1]{{}}
\newcommand{\unicode}[1]{{}}

\usepackage[normalem]{ulem}

\newcommand{\ZANEW}[1]{#1}

\title[Shape of Dwarf From Distance Gradient]{Constraining the shape of Milky Way satellites with distance gradients}

\author[Z. An et al.]{
Zhaozhou An,$^{1}$\thanks{E-mail: za@andrew.cmu.edu}
Sergey E. Koposov$^{2, 3, 1}$
\\
$^{1}$McWilliams Center for Cosmology, Carnegie Mellon University, 5000 Forbes Ave, 15213, \ZANEW{USA}\\
$^{2}$Institute for Astronomy, University of Edinburgh, Royal Observatory, Blackford Hill, Edinburgh EH9 3HJ, UK\\
$^{3}$Institute of Astronomy, University of Cambridge, Madingley Road, Cambridge CB3 0HA, UK\\
}

\date{Accepted XXX. Received YYY; in original form ZZZ}

\pubyear{2021}

\begin{document}
\label{firstpage}
\pagerange{\pageref{firstpage}--\pageref{lastpage}}
\maketitle

\begin{abstract}
    We combine the Dark Energy Camera Legacy Survey (DECaLS) DR8 photometry with \textit{Gaia} photometry to study the 3-D structure of Bootes I, Draco, Ursa Minor, Sextans and Sculptor dwarf galaxies using blue horizontal branch (BHB) stars as distance indicators.  \ZANEW{We construct a new colour-absolute magnitude of BHB stars that we use to measure} the distance gradients  within the body of the dwarf galaxies. We detect a statistically significant non-zero gradient \ZANEW{only} in  Sextans and Sculptor. Through modeling of the gradient and 2-D density of the systems by triaxial Plummer models we find that the distance gradients in both dwarf galaxies are inconsistent with prolate shape, but compatible with oblate or triaxial shapes. In order to explain the observed gradients, \ZANEW{oblate models of Sextans and Sculptor need to have a significant intrinsic ellipticity larger than $0.47$ for Sextans
    and $0.46$ for Sculptor.} The flattened oblate shape may imply a significant anisotropy in velocity distribution in order to be consistent with the lack of significant velocity gradients in these systems.

\end{abstract}



\section{Introduction}

The current cosmological paradigm ($\Lambda$CDM) based on cold dark matter and dark energy has been extremely successful in reproducing a vast variety of observations \citep{1985ApJ...292..371D, 1992MNRAS.258P...1E, 2011MNRAS.413..101G, 2005Natur.435..629S, 2011ApJ...740..102K} particularly at large scales. However $\Lambda$CDM predictions on small scales are still not fully supported by data. The list of problems on small scales is well known, including missing satellites problem  \citep{1999ApJ...522...82K, 1999ApJ...524L..19M}, core-cusp \citep{1994ApJ...427L...1F, 1994Natur.370..629M} and too-big-to-fail \citep{2011MNRAS.415L..40B}. While probing the small scale behaviour of dark matter is possible with a variety of tracers,  one of the best targets to resolve some of these problems and probe the nature of dark matter  are  the dwarf spheroidal galaxies, satellites of the Milky Way \citep{2018PhR...761....1B, Bullock2017,2019ARA&A..57..375S}. While being the most frequent type of galaxy in the universe, they are very dark mater dominated, making them  perfect objects to study dark matter without being much influenced by baryons \citep{karachentseva1985atlas, Mateo1998, Tolstoy2009}.

While twenty years ago the sample of known dwarf spheroidal galaxies contained a handful of galaxies, it has been increasing rapidly over the last years thanks to the arrival of large imaging surveys that enable the discoveries of so-called ultra-faint dwarfs \citep{2007ApJ...656L..13I, 2008ApJ...686L..83B,2016ApJ...832...21H, 2016MNRAS.463..712T, Koposov2015}.

The dwarf spheroidal galaxies have been studied extensively both spectroscopically and photometrically over the years however their exact formation mechanism are still unclear. Specifically we have a poor understanding of how the dwarf galaxies observed now around the Milky Way looked before they were accreted onto the Milky Way and what morphological transformations the dwarf galaxies undergo in the accretion process \citep{2001ApJ...547L.123M,2007Natur.445..738M, 2016ApJ...818..193T}. Due to resolution limits of current numerical simulations we also have a limited knowledge of what sets the detailed properties of of the dwarf galaxies such as luminosity and sizes \citep{2011MNRAS.413..101G, 2012MNRAS.422.1203B, 2017MNRAS.471.3547F}.

One of the most crucial questions in studies of dwarf galaxies is the dark matter distribution inside them as this has implications not only on the origin of dwarf galaxies but also on the nature of dark matter itself \citep{2000PhRvL..85.1158H, 2001ApJ...556...93B}. The dark matter densities been recently constrained through a variety of techniques, such as virial mass estimators \citep{2018MNRAS.481.5073E}, Jeans modeling \citep{2007ApJ...663..948G, 2010MNRAS.408.2364S}, distribution function modeling \citep{2017ApJ...850..116L, 2019MNRAS.484.5453C}, globular cluster kinematics \citep{2009MNRAS.399.1275P, 2012MNRAS.426..601C} and half-light radius mass estimators \citep{2009ApJ...704.1274W, 2010MNRAS.406.1220W}. While these techniques have been successful in extracting dark matter masses and densities, they are often relying on multiple assumptions, such as spherical symmetry or rotational symmetry.

With the improvement of the numerical simulation resolution, work by \citet{2007ApJ...671.1135K, 2010MNRAS.405.1119K, 2014MNRAS.439.2863V, 2015MNRAS.447.1112B} used N-body simulations based on $\Lambda$CDM to show that dark matter subhaloes which are likely to host dwarf galaxies usually have a triaxial shape. \citet{2007ApJ...671.1135K} and \citet{2015MNRAS.447.1112B} also show that tidal stripping will reduce the triaxiality of subhaloes and predict that luminous dSphs with relatively low dark matter content are more spherical than faint dark matter-dominated dwarfs.
\ZANEW{Furthermore in more massive dwarf galaxies the feedback from  star formation can reshape and likely align the dark matter and baryon shapes \citep{governato2012}.  \citet{2016MNRAS.460.4466Z}, \citet{2019MNRAS.485..972T} show the shape of stellar distribution of galaxies is correlated with inner dark matter halo shape, and some papers show that the orientation of inner regions of the dark matter haloes is well aligned with the galaxy shape but the outer regions can be substantially
misaligned \citep{2005ApJ...627..647B, 2011MNRAS.415.2607D, 2015MNRAS.453..721V}. Several studies also show that for self-interacting dark matter, the dark matter shape may follow baryonic shape \citep{2014PhRvL.113b1302K, 2018MNRAS.479..359S}, so studying the 3D shape of a galaxy can potentially help us  test different dark matter models.}

Several previous observational studies looked at distribution of projected axis ratios of galaxies to constrain the distribution of intrinsic 3-D shapes \citep{Sandage1970}. \citet{Merritt1996, Kimm2007, Padilla2008} applied this method to the elliptical galaxies and show that the bright elliptical galaxies have triaxial ellipsoid shape but the faint ones are consistent with oblate shape. A more recent study by \citet{2015MNRAS.450.1409S} infers the intrinsic ellipticity distribution of dwarf galaxies in Andromeda system by assuming galaxies have prolate shapes and \citet{2017MNRAS.472.2670S} infers the distribution of 3-D shapes and alignments for dwarf spheroidal galaxies in the Local Group by using 2-D ellipticities, position angles of major axes and distance moduli.
These studies using the distribution of apparent ellipticities \ZANEW{usually require an assumption about random distribution of galaxy orientations} and a large number of sample galaxies to get good constraints on the distribution of intrinsic shapes. However as the number of observed dwarf galaxies is limited, it is useful to try to constrain the intrinsic shapes of individual dwarf galaxies.

The difficulty of constraining the intrinsic shape for single dwarf galaxy is that we only observe stars in projection, and it's hard to infer an accurate line of sight distance for each star, thus a 3-D spatial distribution for stars will have large uncertainty. In this paper we decide to first infer distance gradients of dwarf galaxies using blue horizontal branch (BHB) stars as tracers, then construct 3-D models based on the distance gradients and 2-D density distribution.

We describe the survey data we used and the how we calculate  distance moduli for BHBs in Section~\ref{sec:data}, we then describe how we model the distance gradient of BHB stars with a mixture model in Section~\ref{sec:distance_gradient}. In Section~\ref{sec:3_d_structure} we  construct a 3-D shape model to distance gradient and 2-D density distribution and present the constraints  on the intrinsic shapes of dwarf galaxies. We discuss our results and potential issues in our method as well as present several checks of our results in Section~\ref{sec:discussion}. The paper is concluded with Section~\ref{sec:summary}.

\section{Data}
\label{sec:data}

This work is based on the photometric data from several surveys, specifically \textit{Gaia} DR2 \footnote{This paper is based on \textit{Gaia} DR2, however we have verified that we obtain similar results if we substitute the DR2 data with \textit{Gaia} EDR3. We have decided to stick with DR2 as the extinction prescription for DR2 from \citet{GaiaCollaboration2018} has not been yet updated for DR3.}\citep{Brown2018, Evans2018}, Dark Energy Survey (DES) DR1 \citep{Abbott2018} and Legacy Survey \citep{Dey2019}. In the next sections we briefly introduce these data in the context of measuring precise distances to BHB stars in dwarf galaxies.

\subsection{\textit{Gaia} photometry}
\label{sec:gaia_phot}
\textit{Gaia} satellite was launched in December 2013 \citep{Prusti2016} and produced the second data release \textit{Gaia} DR2 in Apr 2018. The \textit{Gaia} dataset includes a large set of astrometric measurements for more than a billion sources, but also provides an exquisite space-based  multi-band all-sky photometry in three bands. The broad \textit{Gaia} G band covers the wavelength range from 330 to 1050 nm measured for almost 1.7 billion sources with typical uncertainty of 2 mmag at G=17 mag and 10 mmag at G=20 mag \citep{Brown2018, Evans2018}. The two other \textit{Gaia} photometric bands are the BP and RP that cover respectively the blue  and red wavelength ranges (330 nm to 670 nm and 620 nm to 1050 nm). Due to the fact that the BP and RP photometry is measured by integrating over the dispersed spectra, the BP and RP photometry is significantly less precise than the G photometry with  typical uncertainty of 200 mmag at G = 20.

The accuracy of the \textit{Gaia} G band photometry and small level of systematics gives us possibility to explore the distance gradients in various dwarf galaxies based on photometric data alone.

\subsection{DECaLS and DES photometry}
\label{sec:decals_des}

DECaLS (DECam Legacy Survey) is a pre-imaging survey to the DESI spectroscopic survey, which uses the data collected at the Blanco telescope with the DECam camera \citep{2015AJ....150..150F}. DECaLS data covers the entire South Galactic Cap and the $\delta \leq +34$ regions in the North Galactic Cap. DECaLS can reach magnitude limits of $g=24$ and $r=23.4$ \citep{Dey2019} and the average uncertainty at 20 mag for the g and r band are 6 and 8 mmag.
The DECaLS dataset also includes the sources extracted from data obtained as part of the Dark Energy Survey (DES)\citep{Dey2019, Abbott2018} reduced using \textit{The Tractor} software \citep{2016ascl.soft04008L}, \ZANEW{while the DES uses \textit{PSFEx} \citep{2011ASPC..442..435B} and \textit{SourceExtractor} \citep{1996A&AS..117..393B} to extract sources.}  Since in some areas of the DES footprint the DECaLS photometry is missing, we can rely on the catalogs from DR1 of DES instead.

\subsection{Selection of BHB stars}
\label{sec:target_selection}
To identify the BHB stars in the data we cross match \textit{Gaia} with DECaLS data based on the sky position, and use G, BP and RP band from \texttt{gaia\_source} table and g and r band measurements from DECaLS. The g and r band flux measurements are stored as nanomaggies in DECaLS and they are converted to magnitudes by using $\text{mag}=22.5-2.5 \log_{10}(\text{flux})$.

The magnitude limits we use are ${\tt phot\_g\_mean\_mag}<21$ \citep{Brown2018}, $g < 24.0$,  $r<23.4$ and $z<22.5$ \citep{Dey2019}. We use objects with  ${\tt type}=$ 'PSF' and ${\tt gaia\_pointsource}=$ True in DECaLS data to select stars and ${\tt gaia\_duplicated\_source}=$ False to remove duplicate sources.

The extinction for the \textit{Gaia} G band is calculated by by using Equation 1 and Table 1 from \citet{GaiaCollaboration2018} and the extinction for DECaLS photometry is calculated by using coefficients $A/E_{(B-V)}=3.995, 3.214, 2.165$ for g, r, z band, taken from DECaLS website (\url{https://www.legacysurvey.org/dr8/catalogs/#galactic-extinction-coefficients}) which is computed as in \citet{Schlafly2010}. Since we use SFD dust map \citep{1998ApJ...500..525S}, we also apply 14\% recalibration of SFD which is reported by \citet{2010ApJ...725.1175S} when calculating the extinction for the \textit{Gaia} G band \ZANEW{(but not for the g, r, z bands, as coefficients from \citet{Schlafly2010} include the correction).}

The colour range we used for selection of BHB stars is $-0.3<g-r<0$. We also use a $g-r$, $r-z$ colour-colour boundary based on the Equation 6 of \citet{Li2019} to remove possible blue stragglers contaminants.

\begin{equation}
    \begin{aligned}
        r-z-0.1 \leq & 1.07163(g-r)^5 -1.42272(g-r)^4      \\
                     & +0.69476(g-r)^3 -0.12911(g-r)^2     \\
                     & +0.66993(g-r) -0.11368 \leq r-z+0.1
    \end{aligned}
    \label{equa_cc_cut}
\end{equation}

We change the upper bound from $r-z$ to $r-z+0.1$ and add the lower bound $r-z-0.1$ comparing to the \citet{Li2019} original selection to keep as many as possible BHB stars and remove the quasars. We also apply a selection in absolute magnitude in \textit{Gaia} G band ($0<M_G<1$) and we will describe how to calculate $M_G$ in Section~\ref{sec:distance_modulus}.

The selections described above are used for all the dwarf galaxies except Sculptor. The DECaLS does not have full coverage for Sculptor so we use DES DR1 data instead. And the selection for Sculptor is the same for \textit{Gaia} data part. For the DES data, the magnitude limit is  $g < 24.33$,  $r<24.08$ and $z<22.69$. We use ${\tt EXTENDED\_COADD}<=1$ and $-0.05<{\tt spread\_model\_i}<0.05$ to select stars, and we use ${\tt imaflags\_iso\_[grz]}=0$ and ${\tt flags\_[grz]} < 4$ to select high quality data \citep{Abbott2018}. The extinction calculation, BHB stars colour selection and colour-colour boundary are the same as described above.

Due to the magnitude limit of \textit{Gaia} G band, in this paper we select the dwarf galaxies with BHB stars' G band magnitude less than 21. We will select a circle sky coverage with radius equal to five times half-light radius for each dwarf, and we require that the dwarf galaxy has more than 20 potential BHB stars for us to analyse after applying all the selection described above. We also remove the Sagittarius from the list due to heavy contamination. Using the dwarf galaxies list, dwarf centre, distance moduli and half-light radius from \citet{McConnachie2012}, this leaves us with Bootes I, Draco, Ursa Minor, Sculptor and Sextans.

\subsection{BHBs in the centre}
\label{sec:bhb_in_center}
There is substantial evidence of metallicity gradients or distinct stellar components with different metallicities inside different dwarf galaxies \citep{2008ApJ...681L..13B, 2011ApJ...742...20W, 2019ApJ...870L...8K, 2018MNRAS.480..251C}. It is also known that the magnitude of BHBs stars is  likely metallicity dependent  \citep{2004AJ....127..899S, 2013MNRAS.430.1294F}. Given the lack of certainty in these calibrations, the modeling of BHB magnitude distribution when large metallicity gradients or spreads are present is not feasible. Thus for this paper we decide to not analyse stars within one half-light radius \footnote{\ZANEW{We chose the radius based on the observed larger spread in the centers that we
believe is caused by metallicity effects. We also ran several tests where we varied this masking radius to verify that our results are not too sensitive to it. The details are shown in Appendix~\ref{sec:masking_radius_check}
.}} of the centre to avoid these problems.

\begin{figure}
    \includegraphics[width=\columnwidth]{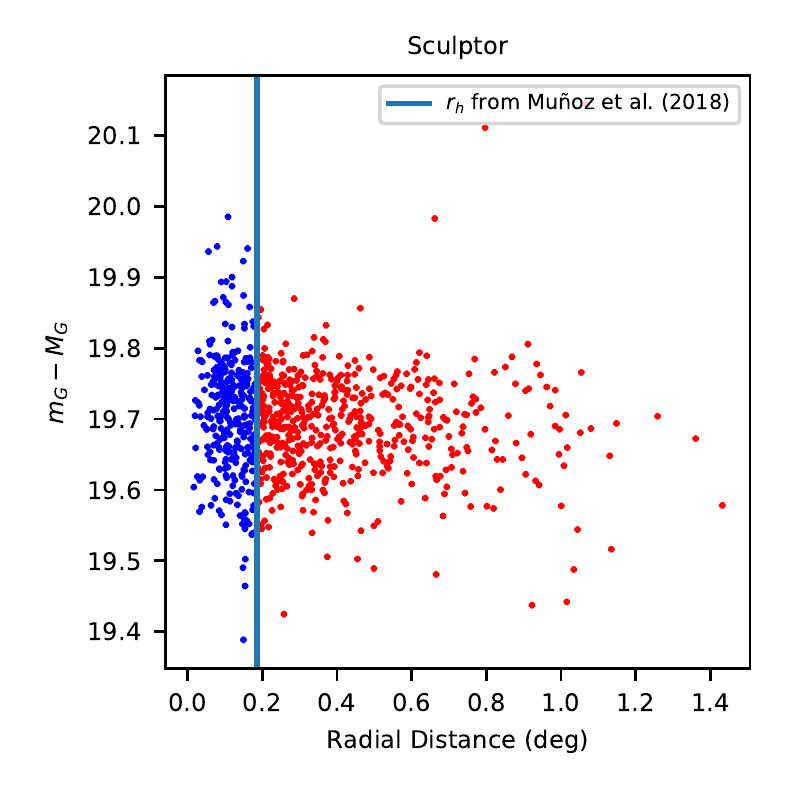}
    \caption{The distance modulus to  BHB stars vs radial distance from the centre in Sculptor dwarf galaxy. The vertical line marks half-light radius from \citet{2018ApJ...860...66M}, we can see the BHB stars which are inside half-light radius have larger dispersion than stars in the outer part. }
    \label{fig_sculptor_gradient_all}
\end{figure}

As an example, Figure~\ref{fig_sculptor_gradient_all} shows the distance modulus ($m_G-M_G$) distribution in radial distance system of BHB stars from Sculptor, where $m_G$ is apparent magnitude in \textit{Gaia} G band and $M_G$ is the absolute magnitude in \textit{Gaia} G band calculated by using the formula in Section~\ref{sec:distance_modulus}. We can see that the stars from inner part have larger distance modulus dispersion than stars from outer part, and our model won't consider the effect of different metallicity components, so we will remove all the blue points in the Figure~\ref{fig_sculptor_gradient_all}.

\subsection{Distance modulus of BHB}
\label{sec:distance_modulus}

With the BHB selection described in previous two sections, the next step to the calculation of distance gradients  is the calculate the distance moduli for individual BHB stars.

BHB stars have almost constant absolute magnitude because of narrow range of masses in the helium burning stage \citep{1970ApJ...161..587I, 1966ApJ...144..995F, 1970A&A.....8..243S}, which makes them excellent distance tracers and have been studied extensively \citep{1983ApJS...53..791P, 1989MNRAS.238..225S, 2002MNRAS.337...87C, 2004AJ....127..899S}. Work by \citet{Deason2011} shows the absolute magnitude slightly depends on the colour. The previous work on colour vs absolute magnitude relation of BHB stars is not based on \textit{Gaia} photometry, so in this paper we determine a new absolute magnitude-color relationship for the \textit{Gaia} G band. And as we discussed in Section~\ref{sec:gaia_phot}, \textit{Gaia} BP and RP bands have a substantial uncertainty at $G\sim$ 20 mag, and most of our target dwarfs have BHB stars which are close to $G\sim20$, so we decide to determine the relation between Gaia $G$ vs DECaLS/DES colour $g-r$.

\begin{figure*}
    \includegraphics[width=\textwidth]{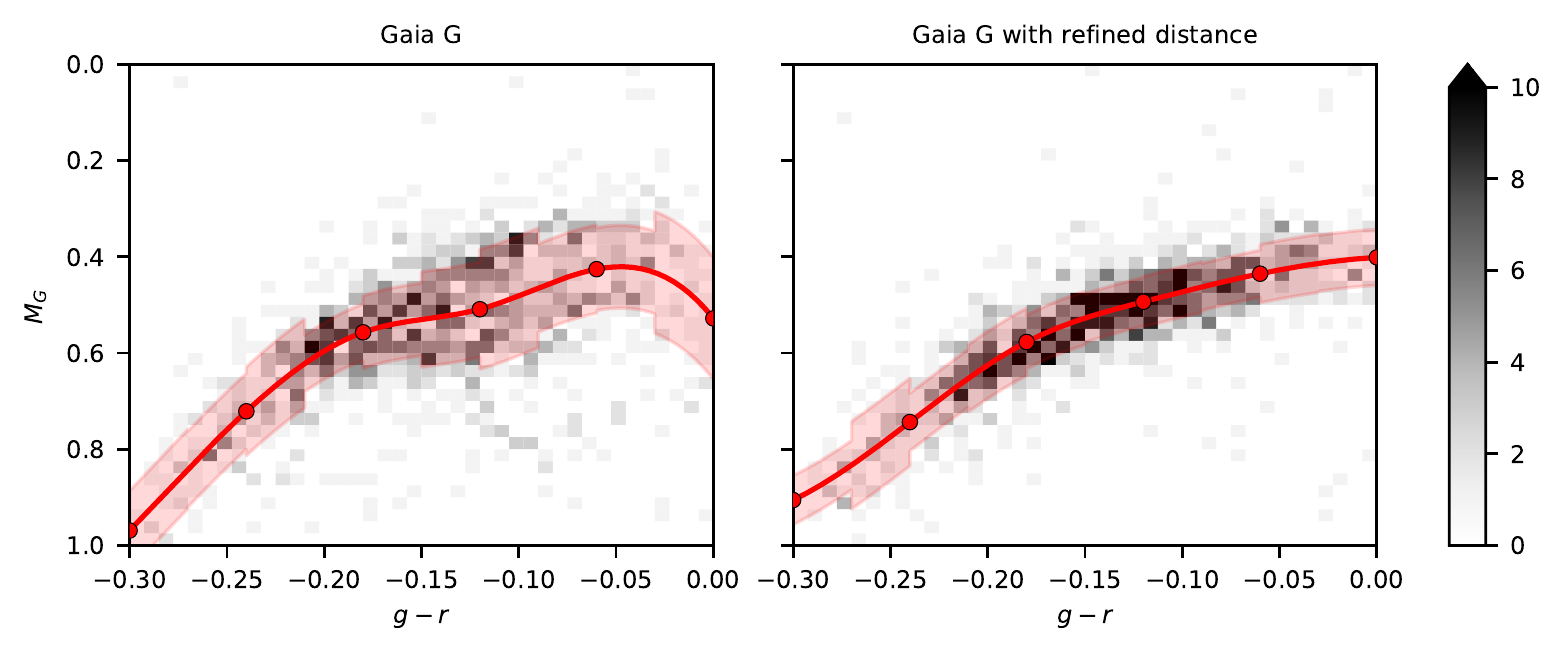}
    \caption{The graph shows the results of fitting BHB colour-absolute magnitude. The colour is DECaLS/DES g-r and the absolute magnitude is calculated from {Gaia} G band. The left panel is using distance moduli from \citep{McConnachie2012} and the right panel is using refined distance moduli which are shown in Table~\ref{tab:distance}. In both panels we show the 2-D histogram in colour-magnitude space for possible BHBs from Bootes I, Draco, Ursa Minor, Sextans and Sculptor. The red line is a best fit spline, the red semi-transparent region shows the fitted intrinsic dispersion; the red dots are the spline knots.}
    \label{fig_abs_mag}
\end{figure*}

To determine the best absolute magnitude vs colour relation for the blue horizontal branch stars we ideally want to use the data for all the dwarf galaxies in our sample.
For this we need to shift the photometric data for each dwarf galaxy by their corresponding distance modulus.  The problem however is that literature distance moduli from \citet{McConnachie2012} are in fact inconsistent with the data we have at hand.
The 2-D histogram in the left panel of Figure~\ref{fig_abs_mag} shows the colour-absolute magnitude distribution of all dwarfs where absolute magnitude is calculated with distance moduli from \citet{McConnachie2012}. We see that the horizontal branch is clearly much thicker than it is expected, indicating incorrect/inconsistent distances in the catalog.
To correct for that we adopt the following iterative procedure. We first use the distances from \citet{McConnachie2012} to fit for the absolute magnitude vs colour relation (left panel of Figure~\ref{fig_abs_mag}). Then we use the fitting result to fit a refined distance modulus for each dwarf separately, and finally use the refined distance moduli to fit the the absolute magnitude vs colour relation again (right panel of Figure~\ref{fig_abs_mag}).

We use six-knot cubic spline with not-a-knot end condition to model absolute magnitude vs colour relation. The BHB data for fitting is selected as we describe in Section~\ref{sec:target_selection} with $M_G$ calculated with corresponding distance moduli data in each step. The whole fitting is done in three steps:
\begin{enumerate}
    \item We use distance moduli from \citet{McConnachie2012}  to calculate $M_G$ in this step. To build the model for fitting, we uniformly divide the colour range into 10 bins and use a mixture model for each bin with a different fraction of member BHB stars, which are assumed to have a Gaussian distribution in $M_G$. The other stars are assumed to have a uniform distribution in $M_G$. We also assume there is an extra uncertainty on $M_G$ which changes with colour bin to accommodate the intrinsic uncertainty of $M_G$ and the uncertainty on the distance modulus. We use maximum likelihood estimation to fit the model and the likelihood is shown below:

          \begin{equation}
              \begin{split}
                  L&= P\left( M_G \mid S, \sigma_i, \alpha_i , g-r\right)\\
                  &=  \alpha_i \mathcal{N}_{trunc}\left( M_{G} \mid f\left(\left(g-r\right) \mid S \right), \sigma_i \right) +(1-\alpha_i)\frac{1}{\Delta M_G}
              \end{split}
              \label{equa_spline_fit}
          \end{equation}
          where $S$ is the set of $M_G$ values of the knots of spline, $i$ is the index of the colour bin the star belongs to, $\sigma_i$ is the intrinsic dispersion for $i$-th colour bin, $\alpha_i$ is the fraction of member BHB stars in the $i$-th colour bin, $\mathcal{N}_{trunc}$ is normal distribution truncated from $0$ to $1$, $f$ is spline function and $\Delta M_G=1$ which is the width of $M_G$ range that we fit. 

    \item Then we refine the distance to each dwarf. We use colour-absolute magnitude relationship from previous steps to fit for distance modulus for each dwarf. When fitting distance modulus, we use the same mixture model as in step (i) except we use a colour independent intrinsic dispersion and add distance modulus parameter for this fitting. We use maximum likelihood estimation and the likelihood is shown below:
          \begin{equation}
              \begin{split}
                  L&= P\left( m_G \mid d ,\sigma, \alpha_i, S, g-r \right)\\
                  &=  \alpha_i \mathcal{N}_{trunc}\left( \left(m_{G}-d\right) \mid f\left(\left(g-r\right) \mid S \right), \sigma \right)\\
                  & \qquad \qquad \quad +(1-\alpha_i)\frac{1}{\Delta M_G}
              \end{split}
              \label{equa_distance_cali}
          \end{equation}
          where $d$ is the distance modulus, $S$ is the set of $M_G$ values of the nodes of spline, $i$ means the colour bin the star belongs to, $\sigma$ is intrinsic dispersion, $\alpha_i$ is the member fraction in the $i$-th colour bin, $m_{G}$ is the apparent magnitude in \textit{Gaia} G band, $\mathcal{N}_{trunc}$ is normal distribution truncated from $0$ to $1$, $f$ is spline function and $\Delta M_G=1$ which is the width of $M_G$ range that we fit. \ZANEW{Unlike Equation~\ref{equa_spline_fit}, we use a colour independent intrinsic dispersion $\sigma$ in this model to reduce the model complexity as we find that the intrinsic dispersions $\sigma_i$ in Equation~\ref{equa_spline_fit} are very close.}
          
    \item \ZANEW{Then we replace the distance moduli in step (i) by the refined distance moduli from step (ii), and keep repeating step (i) and (ii) until convergence. It takes four iterations to get the converged distance moduli. The results are shown in the Table \ref{tab:distance}. We note that the distance moduli determined by our method are determined up to a constant offset common between all the galaxies as we did not calibrate the zero point of the absolute magnitude vs colour curve. This will not affect our analysis we are interested in distance modulus gradients.}

    \item Finally we fit the relation of colour vs absolute magnitude again as in step (i) but now using \ZANEW{the converged distance moduli from step (iii)}.
\end{enumerate}

\begin{table}
    \centering
    \caption{The table for literature distance moduli  $m-M$ from \citet{McConnachie2012} for dwarf galaxies we analysed and our refined distance moduli $(m_G-M_G)_{\text{fit}}$.
    }
    \label{tab:distance}
    \begin{tabular}{ccc} 
        \hline
        Dwarf      & $m-M$ & $(m_G-M_G)_{\text{fit}}$ \\
        \hline
        Bootes I   & 19.11 & \ZANEW{19.129}                 \\
        Draco      & 19.4  & \ZANEW{19.679}                   \\
        Ursa Minor & 19.4  & \ZANEW{19.326}                   \\
        Sculptor   & 19.67 & \ZANEW{19.678}                  \\
        Sextans    & 19.67 & \ZANEW{19.781}                  \\
        \hline
    \end{tabular}
\end{table}

Figure~\ref{fig_abs_mag} shows plots of all these dwarfs' BHB in colour-magnitude diagram and our fitting results. The left panel is the result of step (i) and right panel is the result of step (iv). The average $\sigma_i$ over all the bins is $0.092$ for fitting with distance moduli from \citet{McConnachie2012} and $0.059$ for fitting with refined distance moduli, which demonstrates that improved distance moduli give a  tighter  colour-magnitude relation for BHBs. We can also see that by visually comparing the density distribution in the left panel with the right panel of the Figure~\ref{fig_abs_mag}; the density distribution in the left panel shows larger dispersion than the right panel. The BHB spline track shown on the right panel will be used for calculating distance modulus for each BHB star. The knots of the spline are given in Table~\ref{tab:m_color_knots}. The selection from Section~\ref{sec:target_selection} will use $M_G$ values calculated using refined distance moduli in further analysis.

\begin{table}
    \centering
    \caption{The spline knots for the g-r colour and $M_G$ relationship.
    }
    \label{tab:m_color_knots}
    \begin{tabular}{c|c c c c c c} 
        \hline
        $g-r$       & -0.3   & -0.24  & -0.18  & -0.12  & -0.06  & 0      \\
        \hline
        $M_G$ (mag) & \ZANEW{0.906} &\ZANEW{0.743} &\ZANEW{0.577} &\ZANEW{0.493} &\ZANEW{0.435} &\ZANEW{0.401}  \\
        \hline
    \end{tabular}
\end{table}

\section{The distance gradient of BHBs}
\label{sec:distance_gradient}

Our objective is to constrain the dwarf galaxies 3-D shapes. Here we will assume that the BHB stars in dwarf have the same 3-D spatial shape as the other stars. Under this assumption, we can get the information of a dwarf galaxy shape by modeling the distribution of BHB stars.
As we discussed in Section~\ref{sec:distance_modulus}, BHBs allow us to calculate distance along the line of sight, thus probing not only projected distribution of objects but also the actual 3-D structure.

In this section we will first describe the morphological parameters we use for 2-D spatial distribution of each dwarf galaxy, followed by the description of the model of distance gradient and its measurements for Bootes I, Draco, Ursa Minor, Sextans and Sculptor.

\subsection{The spatial distribution of BHBs}
\label{sec:dwarf_density_dist}
We use a 2-D elliptical Plummer density distribution \citep{1911MNRAS..71..460P} to model the spatial 2-D density distribution for each dwarf galaxy. Given that there are several studies showing that BHB stars can be more spatially extended than other stellar populations \citep{2001MNRAS.327L..15B, 2017MNRAS.467..208O, 2005AJ....130.1065C}, we decided to fit the distribution of BHB stars and check the consistency of half-light radius ($r_h$) between our fitting result and value from \citet{2018ApJ...860...66M}. If two values are consistent we use Plummer parameters from \citet{2018ApJ...860...66M}, otherwise we use half-light radii from our BHB models and all other parameters in Plummer model from \citet{2018ApJ...860...66M} to model density distribution. In this way we can make sure the half-light radius is consistent with data and take advantage of the morphological parameters from \citet{2018ApJ...860...66M} which have small uncertainties.  The model we use is a mixture model of Plummer distribution for member stars and uniform distribution for background, and the likelihood function is given below:

\begin{equation}
    \begin{split}
        L&=P(x, y \mid x_0, y_0, r_h, \theta, \epsilon)\\
        &=\alpha_{mem} \rho(x, y \mid x_0, y_0, r_h, \theta, \epsilon) + \frac{1-\alpha_{mem}}{\Delta S}
    \end{split}
    \label{equa_rh_bhb}
\end{equation}
where $\alpha_{mem}$ is the fraction of member stars, $x, y$ are position in a Cartesian coordinate with units degrees where x direction is along RA and y direction is along Dec ,  $\rho$ is normalized 2-D elliptical Plummer density distribution within the modeled footprint, $x_0, y_0$ are the centre of the dwarf galaxy, $r_h$ is the half-light radius, $\theta$ is the position angle of semi-major axis, $\epsilon$ is the ellipticity and $\Delta S$ is the area of modeled footprint. We use Markov chain Monte Carlo (MCMC) and uniform prior on all parameters to get the posterior mean and $1\sigma$ level uncertainty of the half-light radius for the distribution of BHB stars.

We find only Sextans and Sculptor have posterior mean half-light radii inconsistent with the value from \citet{2018ApJ...860...66M}. The half-light radius in Plummer model from \citet{2018ApJ...860...66M} is $11.17 \pm 0.05$ arcmin for Sculptor and $16.5 \pm 0.10$ arcmin for Sextans, while the half-light radius given by ou r posterior mean is $15.14 \pm 0.67$ arcmin for Sculptor and $43.3 \pm 3.8$ arcmin for Sextans. We note that our half-light radius for BHB stars in Sextans has large difference compared with $27.80 \pm 1.20$ arcmin in \citet{McConnachie2012}, $16.5 \pm 0.10$ arcmin in \citet{2018ApJ...860...66M} and $19.48 \pm 0.35$ arcmin in \citet{2020ApJ...892...27M}, we think the reason is that the BHB stars are much more extended than than other population in Sextans. \citet{2018A&A...609A..53C} also shows the half-light radius for BHB stars in Sextans is $42 \pm 7$ arcmin which is consistent with our result. We decide to use half-light radii from our model for Sextans and Sculptor. The centre of dwarf galaxy, ellipticity and position angle of semi-major axis for Sextans and Sculptor are still taken from \citet{2018ApJ...860...66M}. For other dwarf galaxies all morphological parameters  are taken from \citet{2018ApJ...860...66M}. When masking the centre BHB stars as we said in Section~\ref{sec:bhb_in_center} and performing 3-D model fitting, we always use half-light radius from \citet{2018ApJ...860...66M}.

\subsection{BHB distance gradient model}

With the model of density distribution, we can construct the model for the distance gradient. Since the adopted selection of BHB candidates (see Section~\ref{sec:target_selection}) still includes some foreground contaminant stars on top of BHBs, in order to be able to extract a possible distance gradient we adopt a mixture model for the magnitude distribution.

To model the distribution of BHB's distance moduli, we assume that for member BHBs it linearly depends on the position in the galaxy
\begin{equation}
    \mu_{\mathrm{predict}}=c_{0}+c_{\mathrm{major}} x_{\mathrm{major}}+c_{\mathrm{minor}} x_{\mathrm{minor}}
    \label{equa_dis}
\end{equation}
where $\mu_{\mathrm{predict}}$ is average distance modulus at a given location, $x_{\mathrm{major}}$ and $x_{\mathrm{minor}}$ are coordinates along semi-major and semi-minor in degrees and $c_{\mathrm{major}}$ and $c_{\mathrm{minor}}$ are distance modulus gradients along semi-major/semi-minor axis.
We also assume that the observed distance moduli of individual stars are normally distributed around the prediction as shown below:
\begin{equation}
    \mu \sim \mathcal{N}(\mu_{\mathrm{predict}}, \sqrt{\sigma_{\mathrm{data}}^2+\sigma_0^2}))
    \label{equa_nomral}
\end{equation}
where $\mu$ is the distance modulus of individual BHB star,  $\sigma_{\mathrm{data}}$ is the uncertainty of the observed distance modulus due to photometric errors and $\sigma_0$ is the intrinsic dispersion. The intrinsic dispersion of the calculated BHB absolute magnitudes can be caused by both intrinsic spread of absolute magnitudes of BHBs (i.e. due to age/metallicity spread) or physical distance spread. For the foreground contaminants we assume that distance moduli are uniformly distributed in the selected magnitude range. We note that some foreground contaminants are not necessarily BHBs at correct distance, i.e. they can be more nearby  blue stragglers, however this is not an issue for our model as we are not interested in the true distances to the contaminant population.

To take into account different spatial distribution for  member stars and contaminant stars, we assume Plummer distribution for dwarf galaxy member stars and uniform distribution for contaminant stars (described in Section~\ref{sec:dwarf_density_dist}). By combining the spatial distribution model with magnitude distribution we can write down the likelihood function.

\begin{equation}
    \begin{split}
        L&=P\left(x_{\mathrm{major}}, x_{\mathrm{minor}}, \mu | c_0, c_{\mathrm{major}}, c_{\mathrm{minor}}, \Phi, \sigma_0)\right)\\
        &= \Phi \beta \rho(x_{\mathrm{major}}, x_{\mathrm{minor}}) \mathcal{N}( \mu \mid \mu_{\mathrm{predict}}, \sqrt{\sigma_{\mathrm{data}}^2+\sigma_0^2})  \\
        &\quad +(1-\Phi) \frac{1}{S} \frac{1}{\Delta \mu}
    \end{split}
    \label{equa_pos}
\end{equation}

$\Phi$ is fraction of member stars, $\rho(x_{\mathrm{major}}, x_{\mathrm{minor}})$ is a normalized 2-D elliptical Plummer density distribution within the modeled footprint, $S$ is the area of selected sky coverage, $\Delta \mu$ is the width of selected magnitude range. $\beta$ is normalization factor for the Gaussian given that we only model a finite interval in distance modulus, the equation is shown below,
\begin{equation}
    \beta=1 \left/ \int_{\mu_{min}}^{\mu_{max}} \mathcal{N}(\mu \mid \mu_{\mathrm{predict}}, \sqrt{\sigma_{\mathrm{data}}^2+\sigma_0^2}))  \,d\mu \right.
\end{equation}

\subsection{Gradient fit results}
\label{sec:gradient_fit}
Using the models described in the previous section we can obtain the posterior distribution of distance gradient $(c_{major},c_{minor})$. We use MCMC to sample the posterior while adopting a uniform prior for all the parameters. The priors are specified in the Table~\ref{tab:range_parameters}.

\begin{table*}
    \centering
    \caption{Distance gradient model parameters and their priors. $m_G-M_G$ is the refined distance modulus which is shown in Table~\ref{tab:distance}.}
    \label{tab:range_parameters}
    \begin{tabular}{lcr} 
        \hline
        Parameter                      & Comment                                  & Prior                            \\
        \hline
        $c_0$ (mag)                    & Distance modulus at the centre of galaxy & U[-3+$(m_G-M_G)$, 3+$(m_G-M_G)$] \\
        $c_{\mathrm{major}}$ (mag/deg) & Distance gradient along semi-major axis  & U[-5, 5]                         \\
        $c_{\mathrm{minor}}$ (mag/deg) & Distance gradient along semi-minor axis  & U[-5, 5]                         \\
        $\sigma_0$  (mag)              & Intrinsic distance dispersion            & U[0, 1]                          \\
        $\Phi$                         & Fraction of member stars                 & U[0, 1]                          \\
        \hline
    \end{tabular}
\end{table*}

\begin{table*}
    \centering
    \caption{3-D shape model parameters and their priors.}
    \label{tab:range_parameters_3d}
    \begin{tabular}{lcr} 
        \hline
        Parameter                             & Comment                                                  & Prior              \\
        \hline
        A (deg)                               & Longest principal axis of ellipsoid                      & U[0.1, 1]          \\
        $\epsilon_{3d}$                       & Ellipticity of ellipsoid                                 & U[0, 1]            \\
        $\eta$                                & Triaxiality of ellipsoid                                 & U[0, 1]            \\
        \ZANEW{$\cos \theta_{\text{axis}}$ } & \ZANEW{$\theta_{\text{axis}}$ is a polar angle of the rotation axis}                             & U[-1, 1] \\
        \ZANEW{$\phi$ (rad)}                 & \ZANEW{Azimuthal angle  of the rotation axis} & U[0, 2$\pi$]       \\
        $\psi-\sin \psi$                      & \ZANEW{$\psi$ is a rotation angle}                    & U[0, $\pi$]        \\
        \hline
    \end{tabular}
\end{table*}

\begin{figure*}
    \includegraphics[width=\textwidth]{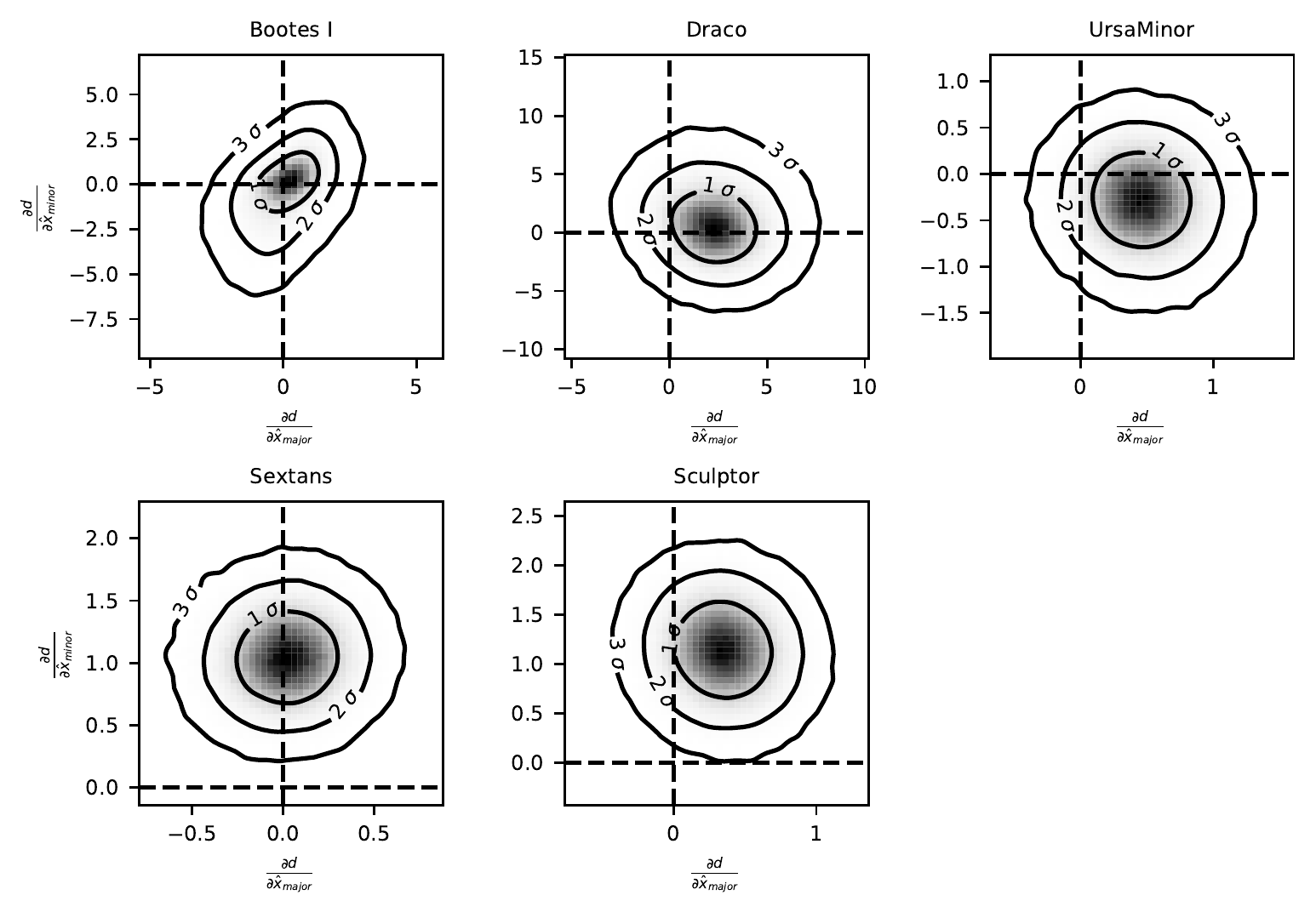}
    \caption{The marginal posterior distribution of distance gradient along semi-major and along semi-minor axes for different dwarf galaxies. $d$ is the distance in pc, $\hat x_{\mathrm{major}}$ and $\hat x_{\mathrm{minor}}$ are the coordinates aligned with the major/minor axes of each dwarf in pc. The zero gradient is marked by the dashed line.}
    \label{fig_gradient_summary}
\end{figure*}

The posterior mean of intrinsic dispersion $\sigma_0$ for these dwarf galaxies is between $0.023$ and $0.064$, which is close to the intrinsic dispersion in the absolute magnitude-colour relationship fitting from Section~\ref{sec:distance_modulus},
and the posterior mean of member fraction $\Phi$ is between $0.72$ and $0.96$ which shows most of stars are from targeted galaxies.
Figure~\ref{fig_gradient_summary} shows marginal posteriors for measured distance gradients in  five dwarf galaxies that we analyze. The zero gradient is marked by dashed lines. Instead of showing the distance modulus gradients in mag/deg as given in  Equation~\ref{equa_dis}, we convert them into dimensionless gradients along semi-major/semi-minor axis.

\begin{equation}
    \begin{split}
        &\hat c_{major} = \frac{\partial d}{\partial \hat{x}_{major}}=\frac{\partial d}{\partial (m_G-M_G)} \frac{\partial x_{major}}{\partial \hat{x}_{major}} \frac{\partial (m_G-M_G)}{\partial  x_{major}}\\
        &\qquad\qquad\qquad\qquad\ =\frac{36  \ln\left(10\right)}{\pi}  \frac{\partial (m_G-M_G)}{\partial  x_{major}} \\
        &\hat c_{minor} = \frac{\partial d}{\partial \hat{x}_{minor}}=\frac{36  \ln\left(10\right)}{\pi}  \frac{\partial (m_G-M_G)}{\partial  x_{major}}
    \end{split}
    \label{equa_conversion_units}
\end{equation}
where $x_{major}, x_{minor}$ are the coordinates on the sky aligned with the major/minor axis expressed in degrees, and $\hat{x}_{major},\hat{x}_{minor}$ are the same coordinates expressed in physical units (pc), $d$ is the line of sight distance in pc. If we assume the galaxy has an oblate or prolate shape, then the dimensionless $\hat c_{major}$ and $\hat c_{minor}$ depend on inclination ($i$), we will have $\hat c_{minor}=\tan(i), \hat c_{major}=0$ for an oblate shape and  $\hat c_{major}=\tan(i), \hat c_{minor}=0$ for a prolate shape.

In the Figures~\ref{fig_gradient_summary}, Sextans and Sculptor show distance gradient at the level above 3 sigma significance. In the case of dwarf galaxies Bootes I, Draco and Ursa Minor, the observed distance gradient is consistent with zero.

Even though we do not observe statistically significant non-zero distance gradient in Bootes I, Draco and Ursa Minor, that is probably because the number of stars is too small and the uncertainty is too large to show any possible distance gradient. A recent paper by \citet{2020MNRAS.tmp.2797M} observes the possible difference of distance between the eastern and western regions of Draco by using RR Lyrae stars, however the error is still too large to get statistically significant conclusion.

We notice that in the case of Sextans and Sculptor the gradient seems to be mostly along the semi-minor axis. We will discuss the implication of this for choosing 3-D model in the next section.

\section{The 3-D structure of galaxy}
\label{sec:3_d_structure}
The results shown in the previous section provide us with the evidence of the distance gradient in two dwarf galaxies. To be able to further investigate the intrinsic 3-D shapes of the Sextans and Sculptor  in this section we will construct a 3-D ellipsoidal model to distance gradient and 2-D density distribution of Sextans and Sculptor.

\subsection{3-D structure model}
\label{sec:3d_model}
We assume the 3-D structure of each dwarf galaxy has a shape of an ellipsoid. As it is often done in the literature to parameterize the ellipsoid we use the 3-D major axis, intrinsic ellipticity and triaxiality. I.e. if we assume the three principal axes of an ellipsoid are $A$, $B$, $C$ with $A \geq B \geq C$, then 3-D major axis is defined as $A$, intrinsic ellipticity (or flattening) is defined as $\epsilon_{3d}=1-\frac{C}{A}$ and triaxiality is defined as $\eta=(1-\frac{B^2}{A^2})/(1-\frac{C^2}{A^2})$ \citep{1991ApJ...383..112F} .

In this paper we focus on 3 families of ellipsoids.
The first one is a general triaxial ellipsoid model that allows arbitrary intrinsic major axis, intrinsic ellipticity and triaxiality. The second family is oblate ellipsoid model (disc-shaped model) which has triaxiality equal to 0, and the last one is prolate ellipsoidal model (cigar-shaped model) which has triaxiality equal to 1.

For each triaxial model we also need to parameterize the orientation of an ellipsoid. This is done by the axis-angle representation, where we parameterize rotation axis by using polar angle $\theta_{\text{axis}}$ and azimuthal angle $\phi$ and we denote rotation angle by $\psi$. The polar angle and azimuthal angle is based on a spherical coordinate system where zenith direction is along line of sight and azimuth reference direction is along RA. We use rotation matrix $R(\theta_{\text{axis}}, \phi, \psi)$ to describe the rotation from the longest principal axis to the RA direction.

With this parameterization and the assumption of the 3-D Plummer profile, we can write the density distribution for our ellipsoid model with principal axes $A,B$ and $C$:

\begin{equation}
    \rho (x, y, z)=\frac{3N}{4\pi} \left(1+X^T R^T S R X\right)^{-\frac{5}{2}} \frac{1}{ABC}
    \label{equa_plummer}
\end{equation}
where $N$ is the total number of stars, $R$ is the rotation matrix described above, $S$ is diagonal matrix with diagonal elements $(1/A^2, 1/B^2, 1/C^2)$, and $X$ is the coordinate in a Cartesian coordinate system where $x$ direction is along RA, $y$ direction is along Dec and $z$ direction is along line of sight.

From the 3-D density distribution, we are able to determine the projected semi-major axis of the stellar distribution ($r_{\text{major}}$), 2-D ellipticity ($\epsilon_{2d}$), positional angle ($\theta_{\text{pos}}$) and the average gradient of the distance along $z$ axis (line of sight direction) across the body of the galaxy.
\begin{equation}
    h(x, y)= \frac{\int_{-\infty}^{+ \infty} \rho(x, y, z) z \,dz}{\int_{-\infty}^{+ \infty} \rho(x, y, z) \,dz}=c_x x+c_y y+c_0
    \label{equa_gradient}
\end{equation}
where   $c_x$ and $c_y$ are average distance gradients along x axis and y axis respectively. We then project the distance gradient along semi-major and semi-minor direction to the x and y direction when doing 3-D model fitting. For the Plummer distribution the calculation of these parameters can be done symbolically. See Appendix~\ref{sec:3d_model_project} for the Mathematica code we use to do this calculation. \ZANEW{We note that the definition of distance gradient we use in Equation~\ref{equa_gradient} assumes the observer is at a infinite distance, while the actual observer has a finite distance to the dwarf galaxy. This difference is negligible in our case.} 

We then model the observed parameters such as 2-D semi-major axis and semi-minor axis, 2-D position angle of semi-major axis and distance gradient using the model described above. We assume that the measured parameters have Gaussian uncertainties. Based on this assumption we construct the likelihood function:

\begin{equation}
    \begin{split}
        L_{3d}&= P\left(\hat r_{\text{major}}, \hat r_{\text{minor}},\hat \theta_{\text{pos}},\hat c_{x},\hat c_{y}|A, \epsilon_{3d}, \eta, \theta_{\text{axis}}, \phi, \psi \right)=\\
        &=\mathcal{N}\left(\hat r_{\text{major}} \mid r_{\text{major}}, \sigma_{r_{\text{major}}}\right) \mathcal{N}\left(\hat r_{\text{minor}} \mid r_{\text{minor}} , \sigma_{r_{\text{minor}}} \right) \\
        &\quad \times \mathcal{N}\left(\hat \theta_{\text{pos}}  \mid \theta_{\text{pos}}, \sigma_{\theta} \right) \mathcal{N}\left(\hat c_{x} \mid c_x , \sigma_{x} \right) \mathcal{N}\left(\hat c_{y} \mid c_y , \sigma_{y}\right)
    \end{split}
    \label{equa_3d_model}
\end{equation}

$ \hat r_{\text{major}},  \hat r_{\text{minor}}, \hat \theta_{\text{pos}}$ are projected semi-major axis, projected semi-minor axis and position angle of dwarf galaxy,  $\hat c_{x} ,\hat c_{y} $ are gradient along x axis and y axis  which is projected from gradient along semi-major and semi-minor axes of dwarf galaxy, all the $\sigma$ are the uncertainty of the corresponding  parameters of dwarf galaxy, and $r_{\text{major}},r_{\text{minor}}, \theta_{\text{pos}}, c_x, c_y$ are corresponding parameters calculated from 3-D model parameterised by $A, \epsilon_{3d}, \eta, \theta_{\text{axis}}, \phi, \psi$.

\subsection{Distance gradient under different 3-D shapes}
\begin{figure*}
\includegraphics[width=\textwidth]{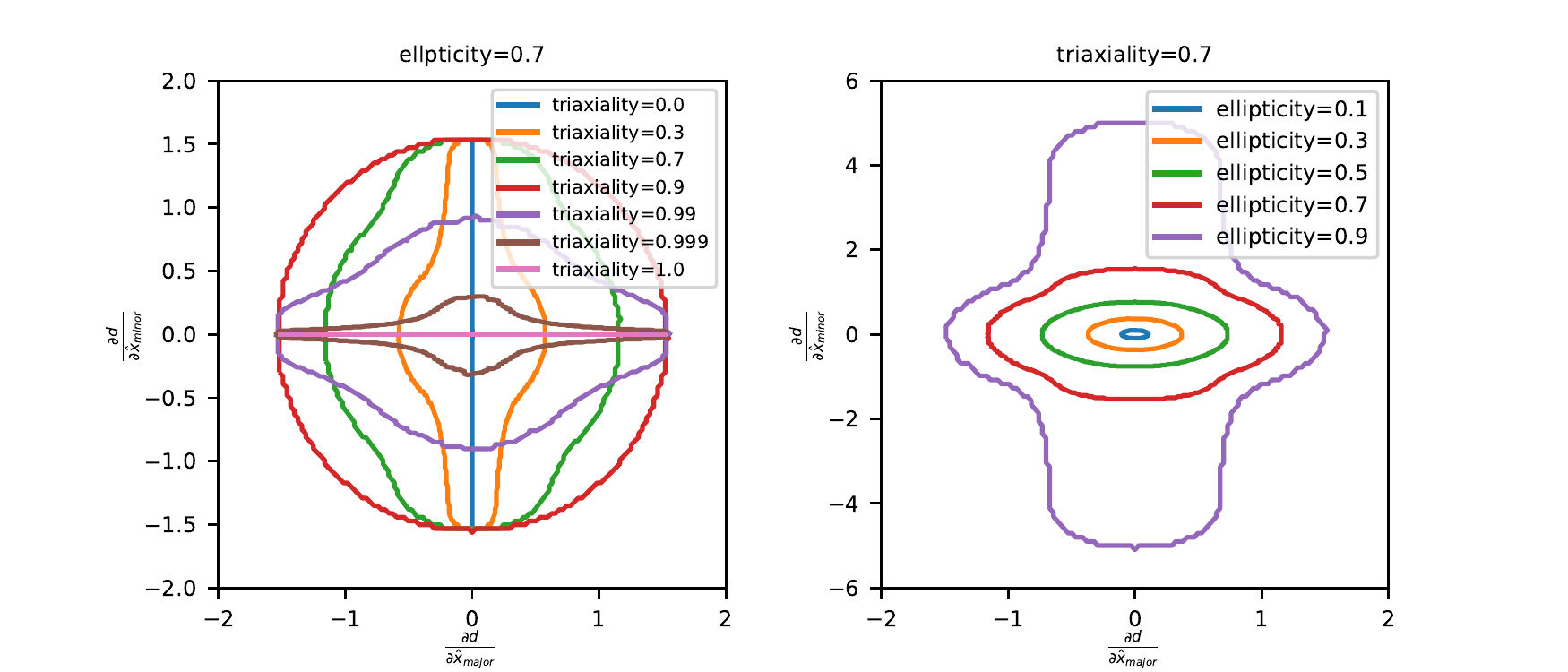}
\caption{\ZANEW{Possible distance gradient values of ellipsoids with different ellipticities and triaxialities along minor  $\frac{\partial d}{\partial x_{min}}$ or major axes $\frac{\partial d}{\partial x_{maj}}$. $d$ is the distance in pc, $\hat x_{\mathrm{major}}$ and $\hat x_{\mathrm{minor}}$ are the coordinates aligned with the major/minor axes in pc. In each panel, different colors represent ellipsoids of different ellipticities and triaxialities. For each ellipsoid the corresponding line shows the boundary of the region of possible distance gradients. I.e. for a triaxial ellipsoid with ellipticity of 0.7 and triaxiality of 0.3 the measured gradients can only be sit within the orange countour depending on the orientation.}}
\label{fig_test_tri_ellip}
\end{figure*}

\ZANEW{Before looking at the results of the modeling we first look at several simulations to build intuition on how ellipticity and triaxiality control the distance gradient. Following the definition of distance gradient in Section~\ref{sec:3d_model}, we calculate the possible distance gradient for shapes with different ellipticity and triaxiality. For each combination of ellipticity and triaxiality, we calculate the distance gradients under all possible rotations, and draw regions of possible distance gradients along two axes. The results are shown in Figure~\ref{fig_test_tri_ellip}. The left panel shows the boundary of the distance gradient distribution when the models have fixed ellipticity $0.7$ but different triaxiality, and the right panel shows the same thing when the models have fixed triaxiality $0.7$ but different ellipticity.  From the left panel we can see that oblate shape ($\text{triaxiality}=0$) has zero gradient along semi-minor axis and prolate shape ($\text{triaxiality}=0$) has zero gradient along semi-major axis, and the triaxiality controls the ratio of the maximum gradient along semi-major axis to the maximum gradient along semi-minor axis. The right panel shows ellipticity controls the maximum magnitude of the gradient. Based on the results from Section~\ref{sec:gradient_fit} where the gradient of Sextans and Sculptor seems to be mostly aligned along the semi-minor axis, the prolate shape is expected to be incompatible with Sextans and Sculptor, so we will only model them with triaxial and oblate shapes.
}

\subsection{3-D shape fit results}
\label{sec:3d_result}
Using the models we described in Section~\ref{sec:3d_model}, we can sample the posterior distribution of each model to infer the the possible intrinsic 3-D shape. Similarly to previous analyses we use MCMC to sample the posterior of our 3-D shape model, and the priors are given in Table~\ref{tab:range_parameters_3d}. The specific prior for \ZANEW{rotation angle $\psi$ from axis-angle representation}   $(\psi-\sin \psi) \sim U[0, \pi]$ is to make sure the sampled rotation matrix from the prior is uniformly distributed with respect to the Haar measure on SO(3) \citep{miles1965random}.

\begin{figure*}
    \includegraphics[width=8cm]{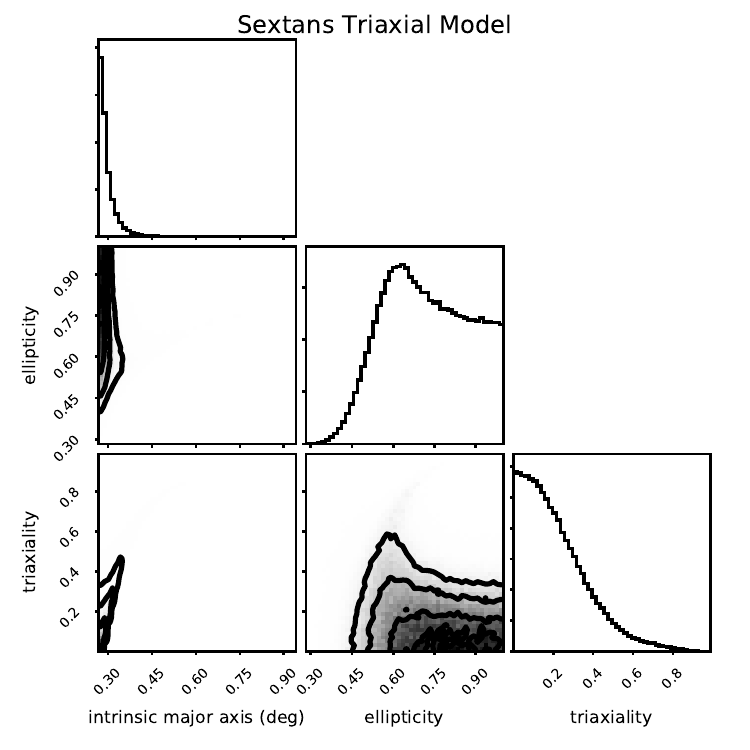}
    \includegraphics[width=8cm]{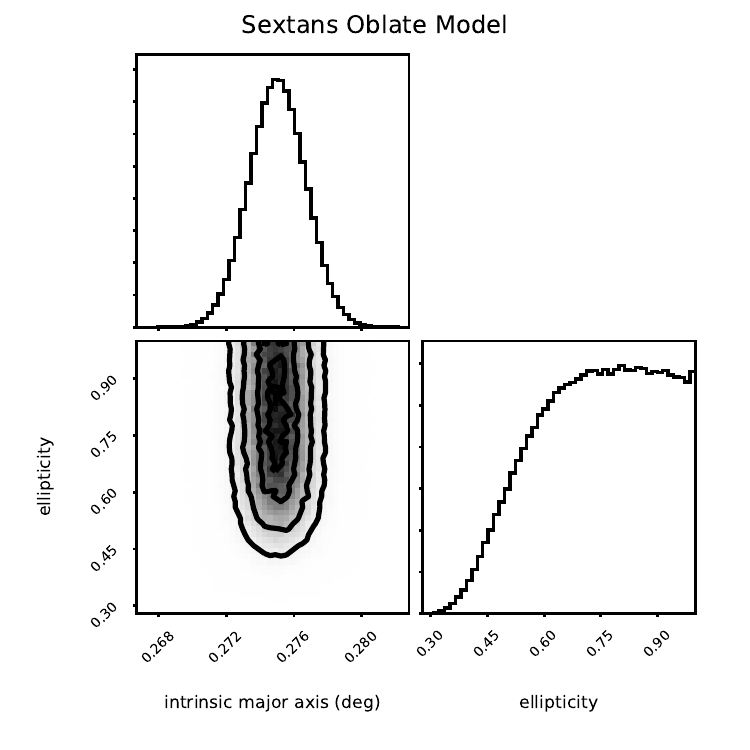}
    \caption{Posterior probability distributions of the parameters of the 3-D model of Sextans. The left panel shows the results for the triaxial model, while the right one shows the oblate model. We do not show here the orientation parameters.
    }
    \label{fig_sextans_3dfit_coner}
\end{figure*}

Figure~\ref{fig_sextans_3dfit_coner} shows the posterior distribution for the parameters of the 3-D model for the oblate and triaxial shapes based on Sextans data. For the triaxial model (left panel), we see that the peak of triaxiality distribution is at zero and the  95\% credible interval for intrinsic ellipticity is $[0.48, 1]$ with peak at 0.6. This says that in order to be consistent with the observed distance gradient and shape the system needs to be quite flattened and of oblate shape. The right panel of the figure shows the posterior for the oblate model, and it provides similar intrinsic ellipticity constraints $[0.47, 1]$.

\begin{figure*}
    \includegraphics[width=8cm]{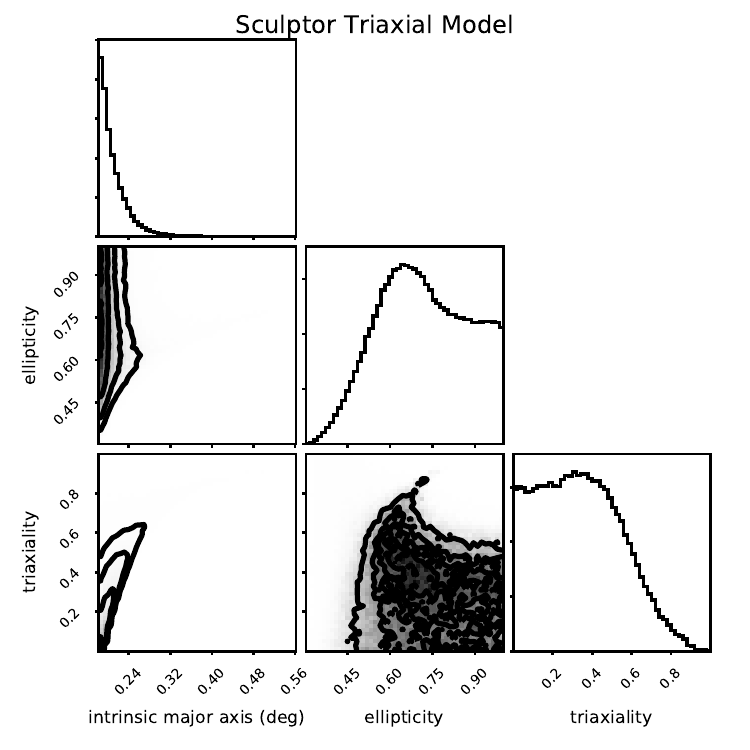}
    \includegraphics[width=8cm]{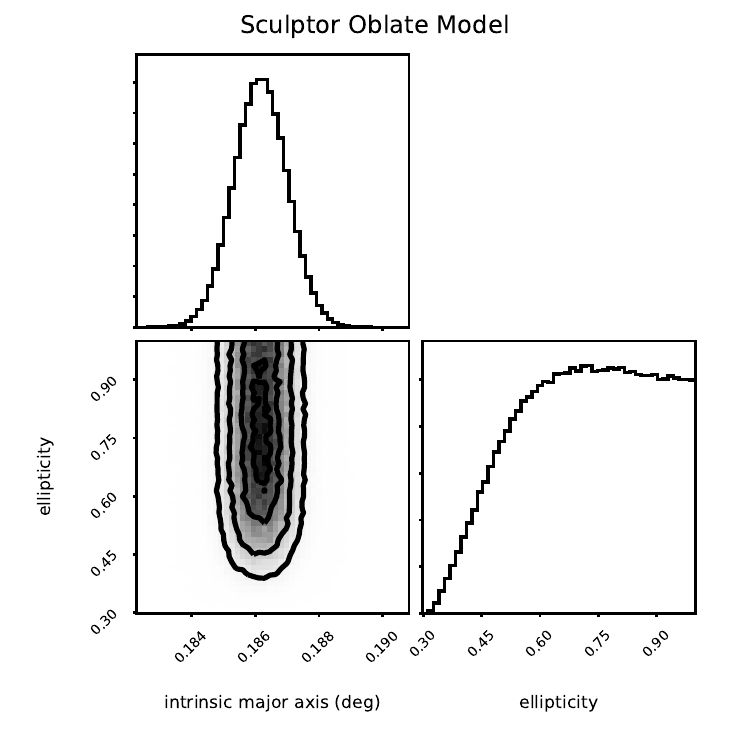}
    \caption{Posterior probability distributions of the parameters of the 3-D model of Sculptor. The left panel shows the results for the triaxial model, while the right one shows the oblate model. The orientation parameters are not shown.}
    \label{fig_sculptor_3dfit_coner}
\end{figure*}

Figure~\ref{fig_sculptor_3dfit_coner} shows the posterior distribution for the parameters of the 3-D model for the oblate and triaxial shapes based on Sculptor data. For the triaxial model (left panel), we see that there is a broad peak in triaxiality distribution at 0.4 and the 95\% credible interval for it is $[0, 0.71]$, which suggests the oblate is possible for Sculptor but the triaxial model doesn't show strong preference for oblate shape. The 95\% credible interval for intrinsic ellipticity is $[0.46, 1]$ with the peak at 0.7. The right panel of the figure shows the posterior for the oblate model, and it provides similar flattening constraints as the triaxial model.

\ZANEW{To validate that the 3-D model can actually fit the measurements we perform posterior predictive checks \citep{gelman2013bayesian} by comparing the measured distance gradient and 2D Plummer parameters to the predicted values based on the samples from the posterior. We check the agreement by computing one tail p-values for each parameter. For Sextans, the one tail p-values are well within $[0.30, 0.5]$ which indicates that the models are in agreement with the data. For Sculptor, the p-values for most parameters are well within $[0.26, 0.5]$ except for the p-value for gradient along semi-major is $0.059$ for oblate model, which indicates the oblate model doesn't fit well to the gradient along semi-major axis. This is understandable since the oblate model is supposed to have exactly zero gradient along semi-major axis, but the last panel of Figure~\ref{fig_gradient_summary} shows the data is consistent with zero gradient along semi-major axis at $1\sigma$ and $2\sigma$ level. Considering the p-value is not extremely small, the oblate model is still acceptable for Sculptor data. The detailed posterior predictive check distributions for Sculptor are shown in the Appendix~\ref{sec:posterior_check}.}

\subsection{Orientation of 3-D shape}
\label{sec:orientation}

\ZANEW{With the posterior of the rotation parameters at hand, we can analyse the orientation of Sextans and Sculptor. Considering that direction of the the major axis of an oblate ellipsoid is not defined and the results of the triaxial model  prefer the oblate shape, we decide to use the direction of minor axis as the orientation. To describe the orientation, we calculate the angle between the minor axis and the velocity of the dwarf ($\theta_{\text{vel}}$) and the angle between the minor axis and the direction from the centre of the dwarf to the Galactic centre ($\theta_{\text{centre}}$). We adopt the peculiar motion of the Sun from \citep{10.1111/j.1365-2966.2010.16253.x}, the Sun's distance to the Galactic centre of $8.3$ kpc \citep{2009ApJ...692.1075G}, and the position of the Galactic centre from \citep{2004ApJ...616..872R}. We use the dwarfs' mean proper motions from \citet{2021arXiv211108737Q} and heliocentric velocity and heliocentric distances from \citet{McConnachie2012}.}


\ZANEW{We sample the rotation parameters from the posterior and the velocity and the distance from the literature values, then calculate $\theta_{\text{vel}}$ and $\theta_{\text{centre}}$.
The mean value and the 1 sigma level asymmetric uncertainties are shown in Table~\ref{tab:angle_result}. We notice the uncertainty of the oblate model is smaller than the uncertainty of the triaxial model, and we think the reason is that oblate model puts a stronger constraint on the possible orientation. The values listed in the Table~\ref{tab:angle_result} do not show clear alignment neither along the velocity of the dwarfs, nor towards the Galactic centre.}


\begin{table}
    \centering
    \caption{Orientation results of Sextans and Sculptor. $\theta_{\text{vel}}$ is the angle between the minor axis and the velocity of the dwarf. $\theta_{\text{centre}}$ is the angle between the minor axis and the direction from the centre of the dwarf to the Galactic centre. We use 1 sigma level asymmetrical uncertainty}
    \label{tab:angle_result}
    \begin{tabular}{cccc}
        \hline
        Galaxy &  Model & $\theta_{\text{vel}}$ &  $\theta_{\text{centre}}$ \\
         & & (degree) & (degree)  \\
        \hline
        Sextans & Oblate & $120.3_{-3.4}^{+3.7}$ & $53.6_{-3.3}^{+3.6}$ \\
        \\
        Sextans & Triaxial & $122.5_{-7.0}^{+7.9}$ & $56.1_{-7.0}^{+7.9}$ \\
        \\
        Sculptor & Oblate & $36.62_{-0.96}^{+0.90}$ & $49.4_{-4.3}^{+4.8}$ \\
        \\
        Sculptor & Triaxial & $43.4_{-6.8}^{+7.0}$ & $51.0_{-8.8}^{+10}$ \\
        \hline
    \end{tabular}
\end{table}


\section{Discussion}
\label{sec:discussion}
In this section we will discuss possible implications of our measurements presented in Section~\ref{sec:3d_result}, potential issues caused by non-axisymmetric metallicity distribution in the galaxies and the possibility of that the shape is the result of tidal disruption. We will discuss the consistency between our measurements and other paper. We also verify that our results are robust to small changes in extinction and the details of this check are shown in the Appendix~\ref{sec:extinction_check}.

\subsection{Intrinsic ellipticity and rotation}

Taken at face value the measurements presented in Section~\ref{sec:3d_result} show that the the systems require significant flattening, which has implications on the dynamical structure the systems. Specifically we expect that if the flattening is strong enough that may be only possible if the system is rotating \citep{Binney2005}. Because the results from Section~\ref{sec:3d_result} show that both Sextans and Sculptor can have oblate shapes, it is possible for us to perform the consistency check between intrinsic ellipticity, anisotropy and rotation (which is constrained by existing observations).

Multiple studies \citep{1994MNRAS.269..957H, 2008ApJ...688L..75W, 2011MNRAS.411.1013B} show there isn't a statistically significant velocity gradient in Sextans, so we need to check whether our intrinsic ellipticity is consistent with the observation of velocity gradient.
From our 3-D model fitting results, we find the intrinsic ellipticity $1-C/A$ for Sextans needs to be above $0.47$, and according to \citet{2008ApJ...688L..75W} the $3\sigma$ upper bound for the absolute value of line of sight velocity gradient is $|-2.1-0.8\times 3|=4.5$ km\,s$^{-1}$\,deg$^{-1}$. \citet{2009AJ....137.3109W} shows the velocity dispersion is $7.9$ km\,s$^{-1}$ in Sextans. To check whether these results are consistent, we can rely on the anisotropic rotating spheroids models presented in \citet{Emsellem2011}, who provides the models for oblate with different tangential anisotropy
and different amount of rotation measured by dimensionless parameter
$\lambda_R$ \citep{2007MNRAS.379..401E} which is defined as $
    \lambda_R \equiv \frac{<R|V|>}{<R\sqrt{V^2+\sigma^2}>}
$
(here $R$ is distance to the centre, $V$ is mean stellar velocity and $\sigma$ is mean stellar velocity dispersion and $<>$ refers to luminosity-weighted sky average).

The flattening of the object can be explained by both rotation and anisotropy of orbits \citep{1976MNRAS.177...19B,1978MNRAS.183..501B}. And for a fixed anisotropy we expect a monotonic relationship between intrinsic ellipticity and rotation. And higher anisotropy can support slower rotation at fixed flattening. With the velocity dispersion and velocity gradient limit of Sextans, we can estimate $\lambda_R$ at half-light radius instead of taking luminosity-weighted sky average, which gives an upper bound $\lambda_R <0.15$. Using the models presented in Appendix B of \citet{Emsellem2011} (see figure B4), we expect that the models with mild anisotropy $\beta\sim 0.4-0.6$ should be able to support flattened spheroids with $\epsilon \sim 0.5$ with small or no amount of rotation $\lambda_R\lesssim 0.2$, which is consistent with our estimation. Thus at the face value the flattened shape can be consistent with observed lack of rotation. \ZANEW{We note that previous relation assumes that the dark matter density distribution follows the stellar density and the galaxy is axisymmetric\citep{2006MNRAS.366.1126C}.}

We did the same analysis for oblate shape fitting result of Sculptor. We use $3\sigma$ upper bound for the absolute value of line of sight velocity gradient $|-5.5-0.5\times 3|=7$ km\,s$^{-1}$\,deg$^{-1}$ \citep{2008ApJ...688L..75W} and velocity dispersion $9.2$  km\,s$^{-1}$ \citep{2009AJ....137.3109W} to estimate an upper bound  $\lambda_R <0.14$. The lower bound of intrinsic ellipticity from Section~\ref{sec:3d_result} is $0.46$. We expect that the models with mild anisotropy $\beta\sim 0.4-0.6$ should be able to support flattened spheroids with $\epsilon \sim 0.5$ with small or no amount of rotation $\lambda_R\lesssim 0.2$, which is also consistent with our estimation.

\subsection{Non-axisymmetric metallicity distribution in Sextans}
\label{sec:metallicity}

Our modeling assumes that the stellar distribution is described by ellipsoids and the metallicity distribution is axisymmetric. If the galaxies have substructures with drastically different metallicities, these can affect BHB absolute magnitude calculation, and can mimick themselves as structures at different distances or distance gradients in our model.

One of two systems in which we have detected a significant gradient is Sextans, whose structure has been studied extensively and several possible non-axisymmetric substructures were detected \citep{2019ApJ...870L...8K, 2018MNRAS.480..251C}.

According to the Figure~3 of \citet{2019ApJ...870L...8K} and Figure~5 of \citet{2018MNRAS.480..251C}, there are different metallicity components near the galaxy centre of Sextans. However the sky coverage in these two papers is not large enough and the different metallicity component which is shown in their figures is removed when we remove all the stars inside $1r_h$ as discussed in Section~\ref{sec:bhb_in_center}. Due to limited sky coverage, it is hard to conclude whether there is non-axisymmetric metallicity distribution in the BHB stars we used for the distance gradient fitting.

\begin{figure}
    \includegraphics[width=\columnwidth]{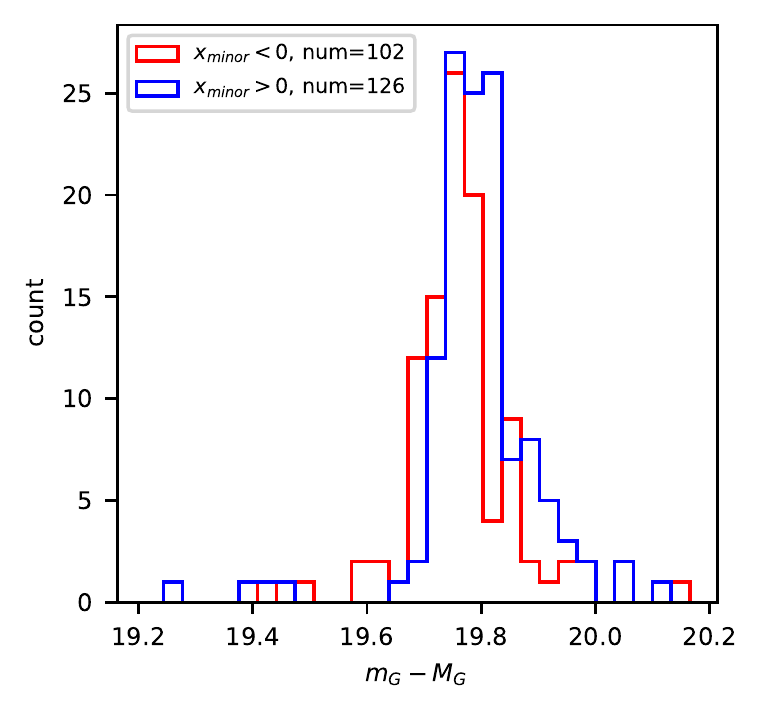}
    \caption{The plot shows the distance modulus distribution of Sextans' BHB stars for different sides separated by semi-minor axis. $x_{minor}$ is the position along semi-minor axis, and the centre of Sextans has $x_{minor}=0$. $x_{minor}>0$ is northwest side and $x_{minor}<0$ is southeast side}
    \label{fig_distance_two_parts}
\end{figure}

We also notice that the numbers of BHB stars at different sides of Sextans semi-major axis are somewhat different. Figure~\ref{fig_distance_two_parts} shows the histogram of distance modulus for two galaxy sides ($x_{minor}>0$ is northwest side and $x_{minor}<0$ is southeast side) without removing the stars within $1r_h$ of Sextans, and the number of BHB stars at northwest side is larger than the other side by a factor of $1.24$. The side with more BHB stars is the same side where \citet{2019ApJ...870L...8K} claims there is a overdensity of metal-poor stars. Due to the limit of their sky coverage and the small number of BHB stars in the centre region, we cannot get a reliable conclusion about whether the this overdensity is related with the difference of number of BHB stars between different side of semi-major axis. We did a similar analysis of the BHB spatial distribution in Sculptor and we do not find any significant difference between the different sides of the axes.

The non-axisymmetric metallicity distribution in Sextans is a potential issue that may affect our distance calculation, but with current data we cannot know how much it affects. Future data with larger sky coverage may help to resolve this.

\subsection{Tidal disruption}
One of the possible explanations of the flattening and elongation of dwarf galaxies is tidal disruption. The most famous example around the Milky Way is the  Sagittarius dwarf galaxy. Sagittarius core which is undergoing strong tidal disruption shows a high ellipticity $\epsilon=0.62-0.65$ \citep{2003ApJ...599.1082M}. For Sextans and Sculptor, the orbits of both galaxies imply that they are not likely to be affected by tides \citep{2019MNRAS.490..231K}, and \citet{2016MNRAS.460...30R} performs structural analysis and conclude that Sextans is not undergoing significant tidal disruption from the Milky Way, however a recent study by \citet{2019AJ....157...35V} found two RR Lyrae stars and one anomalous Cepheid that may be extratidal stars of Sextans, which suggest that this galaxy may be undergoing tidal destruction. For Sculptor, \citet{2006AJ....131..375W} states the possibility of past tidal disruption could have occurred given that there is possible presence of unbound tidal debris and \citet{2009ApJ...698..222P} also states the possibility of past tidal disruption considering the shape of its density distribution, but a more recent study by \citet{2011A&A...528A.119D} shows that the clear radial gradients in Sculptor means lack of signs of recent tidal interactions. \ZANEW{Our shape modeling results also prefer oblate shapes and do not show alignment of the best-fit ellipsoids towards the Galactic centre, while the numerical simulations suggest that tidally stripped  subhaloes should have prolate shapes oriented towards the Galactic centre \citep{2010MNRAS.406..744C, 2014MNRAS.439.2863V, 2015MNRAS.447.1112B}}.
With all this evidence, we think it is unlikely that the observed 3-D shape is caused by tidal disruption.

\subsection{Consistency with other shape studies}
Our results for Sextans and Sculptor are consistent with results from \citet{2017MNRAS.472.2670S}, where the authors measured the intrinsic shapes and alignments of dSph galaxies of Local Group and find the dSph of the Milky Way have mean intrinsic ellipticities around 0.6. There is also a recent work by \citet{2020ApJ...900...69X} shows that faint Local Group dwarf galaxies are more likely to be oblate while the bright one are more likely to be prolate based on the correlation between 2D ellipticity and central surface brightness, and the mass-to-light threshold between bright and faint is between $70-200 M_{\sun}/L_{\sun}$. Our result for Sextans which has mass-to-light $98 M_{\sun}/L_{\sun}$ is consistent with their conclusion, however the distance gradient of Sculptor which has mass-to-light $12 M_{\sun}/L_{\sun}$ is not consistent with prolate shape. \citet{2020ApJ...900...69X} doesn't reject the possibility that the shape could be triaxial, and our result shows that Sculptor possibly have a non-zero triaxiality.

\section{Conclusions}
\label{sec:summary}

Using the data from \textit{Gaia} DR2, DECaLS and DES, we perform the modeling of possible distance gradient in five dwarf galaxies using BHB stars. The results from Bootes I, Draco and Ursa Minor are all consistent with zero gradient within $3\sigma$, but Sextans and Sculptor show statistically significant non-zero distance gradient. In both of these cases the distance gradient is along semi-minor axis, with  the distance gradient along semi-major axis  consistent with zero.

We construct 3-D ellipsoid models for Sextans and Sculptor to explain the distance gradient. The results show that an oblate shape is preferred for Sextans and intrinsic ellipticity ($1-C/A$) larger than $0.47$. For Sculptor the result of triaxial model shows possible oblate shape with intrinsic ellipticity larger than $0.46$ and possibly non-zero triaxiality.

We explore the validity of the oblate models for Sextans and Sculptor by checking the relationship between galaxy rotation, anisotropy and intrinsic ellipticity, we find our results are consistent with current constraints on the velocity gradient in these systems with mild anisotropy.

One potential issue that could bias our detections is the presence of non-axisymmetric structures with different metallicities, such as have been previously seen in Sextans. We also see a different number of BHB stars at different sides of the galaxy that could be related with this, however we cannot get a reliable conclusion with current data.

We show that with our method it is possible to constrain the 3-D shape of individual dwarf galaxies. With more systems and other distance indicators such as RR Lyrae, red clump stars we can hope to obtain better constraints on the  3-D shapes of dwarf galaxies.

\section{ACKNOWLEDGEMENTS}

This paper made use of the Whole Sky Database (wsdb) created by Sergey Koposov and maintained at the Institute of Astronomy, Cambridge by Sergey Koposov, Vasily Belokurov and Wyn Evans with financial support from the Science \& Technology Facilities Council (STFC) and the European Research Council (ERC).

This software made use of the Q3C software \citep{Koposov2006}, NumPy \citep{harris2020array}, SciPy \citep{2020SciPy-NMeth}, Matplotlib \citep{4160265}, emcee \citep{emcee}, Mathematica \citep{Mathematica}, Numba \citep{10.1145/2833157.2833162}.

This work has made use of data from the European Space Agency (ESA) mission
    {\it Gaia} (\url{https://www.cosmos.esa.int/gaia}), processed by the {\it Gaia}
Data Processing and Analysis Consortium (DPAC,
\url{https://www.cosmos.esa.int/web/gaia/dpac/consortium}). Funding for the DPAC
has been provided by national institutions, in particular the institutions
participating in the {\it Gaia} Multilateral Agreement.

The Legacy Surveys consist of three individual and complementary projects: the Dark Energy Camera Legacy Survey (DECaLS; NOAO Proposal ID \# 2014B-0404; PIs: David Schlegel and Arjun Dey), the Beijing-Arizona Sky Survey (BASS; NOAO Proposal ID \# 2015A-0801; PIs: Zhou Xu and Xiaohui Fan), and the Mayall z-band Legacy Survey (MzLS; NOAO Proposal ID \# 2016A-0453; PI: Arjun Dey). DECaLS, BASS and MzLS together include data obtained, respectively, at the Blanco telescope, Cerro Tololo Inter-American Observatory, National Optical Astronomy Observatory (NOAO); the Bok telescope, Steward Observatory, University of Arizona; and the Mayall telescope, Kitt Peak National Observatory, NOAO. The Legacy Surveys project is honored to be permitted to conduct astronomical research on Iolkam Du'ag (Kitt Peak), a mountain with particular significance to the Tohono O'odham Nation.

NOAO is operated by the Association of Universities for Research in Astronomy (AURA) under a cooperative agreement with the National Science Foundation.

This project used data obtained with the Dark Energy Camera (DECam), which was constructed by the Dark Energy Survey (DES) collaboration. Funding for the DES Projects has been provided by the U.S. Department of Energy, the U.S. National Science Foundation, the Ministry of Science and Education of Spain, the Science and Technology Facilities Council of the United Kingdom, the Higher Education Funding Council for England, the National Center for Supercomputing Applications at the University of Illinois at Urbana-Champaign, the Kavli Institute of Cosmological Physics at the University of Chicago, Center for Cosmology and Astro-Particle Physics at the Ohio State University, the Mitchell Institute for Fundamental Physics and Astronomy at Texas A\&M University, Financiadora de Estudos e Projetos, Fundacao Carlos Chagas Filho de Amparo, Financiadora de Estudos e Projetos, Fundacao Carlos Chagas Filho de Amparo a Pesquisa do Estado do Rio de Janeiro, Conselho Nacional de Desenvolvimento Cientifico e Tecnologico and the Ministerio da Ciencia, Tecnologia e Inovacao, the Deutsche Forschungsgemeinschaft and the Collaborating Institutions in the Dark Energy Survey. The Collaborating Institutions are Argonne National Laboratory, the University of California at Santa Cruz, the University of Cambridge, Centro de Investigaciones Energeticas, Medioambientales y Tecnologicas-Madrid, the University of Chicago, University College London, the DES-Brazil Consortium, the University of Edinburgh, the Eidgenossische Technische Hochschule (ETH) Zurich, Fermi National Accelerator Laboratory, the University of Illinois at Urbana-Champaign, the Institut de Ciencies de l'Espai (IEEC/CSIC), the Institut de Fisica d'Altes Energies, Lawrence Berkeley National Laboratory, the Ludwig-Maximilians Universitat Munchen and the associated Excellence Cluster Universe, the University of Michigan, the National Optical Astronomy Observatory, the University of Nottingham, the Ohio State University, the University of Pennsylvania, the University of Portsmouth, SLAC National Accelerator Laboratory, Stanford University, the University of Sussex, and Texas A\&M University.

BASS is a key project of the Telescope Access Program (TAP), which has been funded by the National Astronomical Observatories of China, the Chinese Academy of Sciences (the Strategic Priority Research Program "The Emergence of Cosmological Structures" Grant \# XDB09000000), and the Special Fund for Astronomy from the Ministry of Finance. The BASS is also supported by the External Cooperation Program of Chinese Academy of Sciences (Grant \# 114A11KYSB20160057), and Chinese National Natural Science Foundation (Grant \# 11433005).

The Legacy Survey team makes use of data products from the Near-Earth Object Wide-field Infrared Survey Explorer (NEOWISE), which is a project of the Jet Propulsion Laboratory/California Institute of Technology. NEOWISE is funded by the National Aeronautics and Space Administration.

The Legacy Surveys imaging of the DESI footprint is supported by the Director, Office of Science, Office of High Energy Physics of the U.S. Department of Energy under Contract No. DE-AC02-05CH1123, by the National Energy Research Scientific Computing Center, a DOE Office of Science User Facility under the same contract; and by the U.S. National Science Foundation, Division of Astronomical Sciences under Contract No. AST-0950945 to NOAO.

This project used public archival data from the Dark Energy Survey (DES). Funding for the DES Projects has been provided by the U.S. Department of Energy, the U.S. National Science Foundation, the Ministry of Science and Education of Spain, the Science and Technology FacilitiesCouncil of the United Kingdom, the Higher Education Funding Council for England, the National Center for Supercomputing Applications at the University of Illinois at Urbana-Champaign, the Kavli Institute of Cosmological Physics at the University of Chicago, the Center for Cosmology and Astro-Particle Physics at the Ohio State University, the Mitchell Institute for Fundamental Physics and Astronomy at Texas A\&M University, Financiadora de Estudos e Projetos, Funda{\c c}{\~a}o Carlos Chagas Filho de Amparo {\`a} Pesquisa do Estado do Rio de Janeiro, Conselho Nacional de Desenvolvimento Cient{\'i}fico e Tecnol{\'o}gico and the Minist{\'e}rio da Ci{\^e}ncia, Tecnologia e Inova{\c c}{\~a}o, the Deutsche Forschungsgemeinschaft, and the Collaborating Institutions in the Dark Energy Survey.
The Collaborating Institutions are Argonne National Laboratory, the University of California at Santa Cruz, the University of Cambridge, Centro de Investigaciones Energ{\'e}ticas, Medioambientales y Tecnol{\'o}gicas-Madrid, the University of Chicago, University College London, the DES-Brazil Consortium, the University of Edinburgh, the Eidgen{\"o}ssische Technische Hochschule (ETH) Z{\"u}rich,  Fermi National Accelerator Laboratory, the University of Illinois at Urbana-Champaign, the Institut de Ci{\`e}ncies de l'Espai (IEEC/CSIC), the Institut de F{\'i}sica d'Altes Energies, Lawrence Berkeley National Laboratory, the Ludwig-Maximilians Universit{\"a}t M{\"u}nchen and the associated Excellence Cluster Universe, the University of Michigan, the National Optical Astronomy Observatory, the University of Nottingham, The Ohio State University, the OzDES Membership Consortium, the University of Pennsylvania, the University of Portsmouth, SLAC National Accelerator Laboratory, Stanford University, the University of Sussex, and Texas A\&M University.
Based in part on observations at Cerro Tololo Inter-American Observatory, National Optical Astronomy Observatory, which is operated by the Association of Universities for Research in Astronomy (AURA) under a cooperative agreement with the National Science Foundation.

\section*{Data Availability}
The data underlying this article are derived from sources in the public domain: \url{https://www.cosmos.esa.int/web/gaia/dr2}, \url{https://www.legacysurvey.org/dr8/} and \url{https://www.darkenergysurvey.org/dr1-data-release-papers/}.

\bibliographystyle{mnras}
\bibliography{ref}

\appendix

\section{Masking of central regions}
\label{sec:masking_radius_check}
\begin{figure*}
    \includegraphics[width=\textwidth]{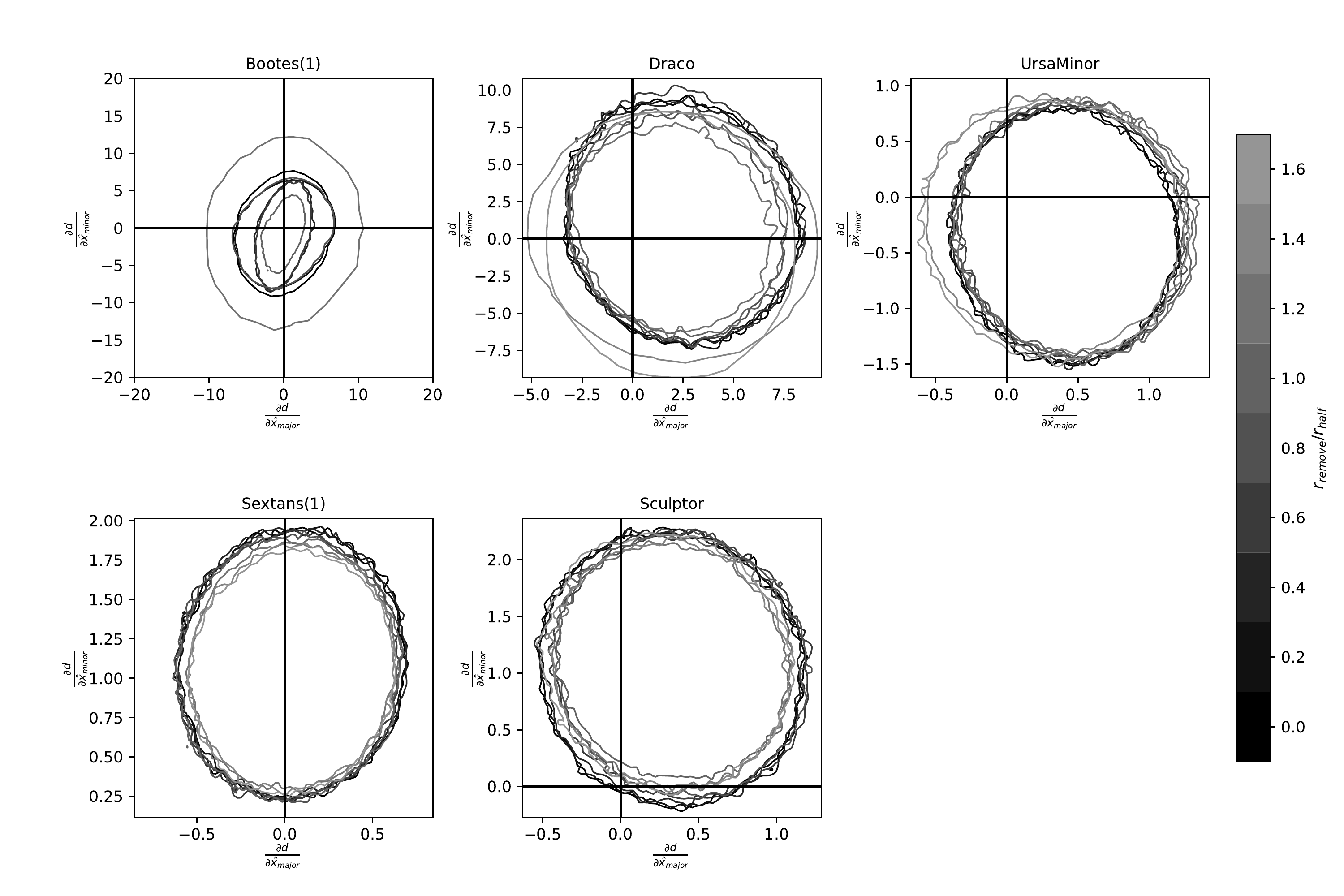}
    \caption{The posterior of distance gradient with different masking radius. $d$ is the distance in pc. $\hat x_{\mathrm{major}}$ and $\hat x_{\mathrm{minor}}$ are the coordinates aligned with the major/minor axes of each dwarf in pc. The zero gradient is marked by the dashed line. Each contour line represents a $3\sigma$ boundary of distance gradient posterior with a specific masking radius; the ratio of masking radius to half light radius is coded by the grayscale of the contour.}
    \label{fig_removal_radius_test}
\end{figure*}

\ZANEW{In Figure \ref{fig_sculptor_gradient_all} we showed the distance modulus versus radial distance of Sculptor, which suggested the masking of stars within one half light radius. While this might seem reasonable for Sculptor based on this plot, it might not be suitable for other dwarfs. To further check how sensitive our
results are to the masking radius, we run our distance gradient analysis with different masking radius. The distance gradient results are shown in Figure \ref{fig_removal_radius_test}. For most dwarfs, the change of the distance gradient distribution is small and negligible and our conclusion remains the same. The large change of Bootes I is due to the small number of BHB stars, so the masking of inner part will dramatically decrease the total number of BHB and leads to the large change in the posterior. And for Sculptor, we notice when we do not mask stars at all or only mask stars within a small radius, we do not have 3 sigma level significance of non-zero gradient, but when we remove with a radius no less than $0.6r_h$ we find the non-zero gradient is significant. Considering Figure \ref{fig_sculptor_gradient_all} shows the Sculptor has large distance dispersion in the inner part, we think it is reasonable to mask the inner part, and our results are not too sensitive to the choice of the masking radius.}

\section{3-D Model Projection Code}
\label{sec:3d_model_project}
First we define parameters for triaxial ellipsoid. Here we use Euler angles to parameterize rotation, the parameterization will be converted to axis-angle representation when we perform 3-D shape fitting.

\begin{verbatim}
Rx[a_]:=RotationMatrix[a, {1, 0, 0}]
Ry[b_]:=RotationMatrix[b, {0, 1, 0}]
Rz[c_]:=RotationMatrix[c, {0, 0, 1}]
B=DiagonalMatrix[{1/mx^2,1/my^2,1/mz^2}]
\end{verbatim}
Then we integrate over $z$ to get 2-D Plummer distribution. We will use $m, n, p$ to represent the factors which are independent of $z$ in integral. \verb|EllipExpression| is quadratic function of $x$ and $y$ which is extracted from the result of previous integral, and it will be used for calculating the shape of 2-D distribution.

\begin{verbatim}
CoeffZ=CoefficientList[
    1+X.Transpose[Rx[a]].Transpose[Ry[b]].
    Transpose[Rz[c]].B.Rz[c].Ry[b].Rx[a].X, z]
Integrate[1/(m+n*z+p*z^2)^(5/2), 
    {z, -\[Infinity], \[Infinity]}, Assumptions->
    m\[Element]Reals&&n\[Element]Reals&&
    p\[Element]Reals&&n^2-4 m p<0&&m>0
EllipExpression=n^2-4 m p/.{m->CoeffZ[[1]],
    n->CoeffZ[[2]], p->CoeffZ[[3]]}
\end{verbatim}
Then we calculate position angle of semi-major axis vector and ratio of semi-major to semi-minor. The calculation is done by finding the eigenvalues and eigenvectors of the symmetric matrix of quadratic function of $x, y$. The values in \verb|Evalues| are semi-minor and semi-major axis, and the corresponding column in the \verb|Evectors| represents the vector along semi-minor/semi-major axis.

\begin{verbatim}
CoeffXY=CoefficientList[EllipExpression, {x, y}]
EllipCoefMatrix={{D1, D2/2}, {D2/2,D3}}
Evalues=Eigenvalues[EllipCoefMatrix]
Evectors=Eigenvectors[EllipCoefMatrix]
Evalues[[2]]/.{D1->CoeffXY[[3, 1]]/CoeffXY[[1, 1]],
    D2->CoeffXY[[2, 2]]/CoeffXY[[1, 1]],
    D3->CoeffXY[[1, 3]]/CoeffXY[[1, 1]]}
Evectors/.{D1->CoeffXY[[3, 1]]/CoeffXY[[1, 1]],
    D2->CoeffXY[[2, 2]]/CoeffXY[[1, 1]],
    D3->CoeffXY[[1, 3]]/CoeffXY[[1, 1]]}
\end{verbatim}

\section{Posterior Predictive Check}
\label{sec:posterior_check}

\begin{figure*}
    \includegraphics[width=\textwidth]{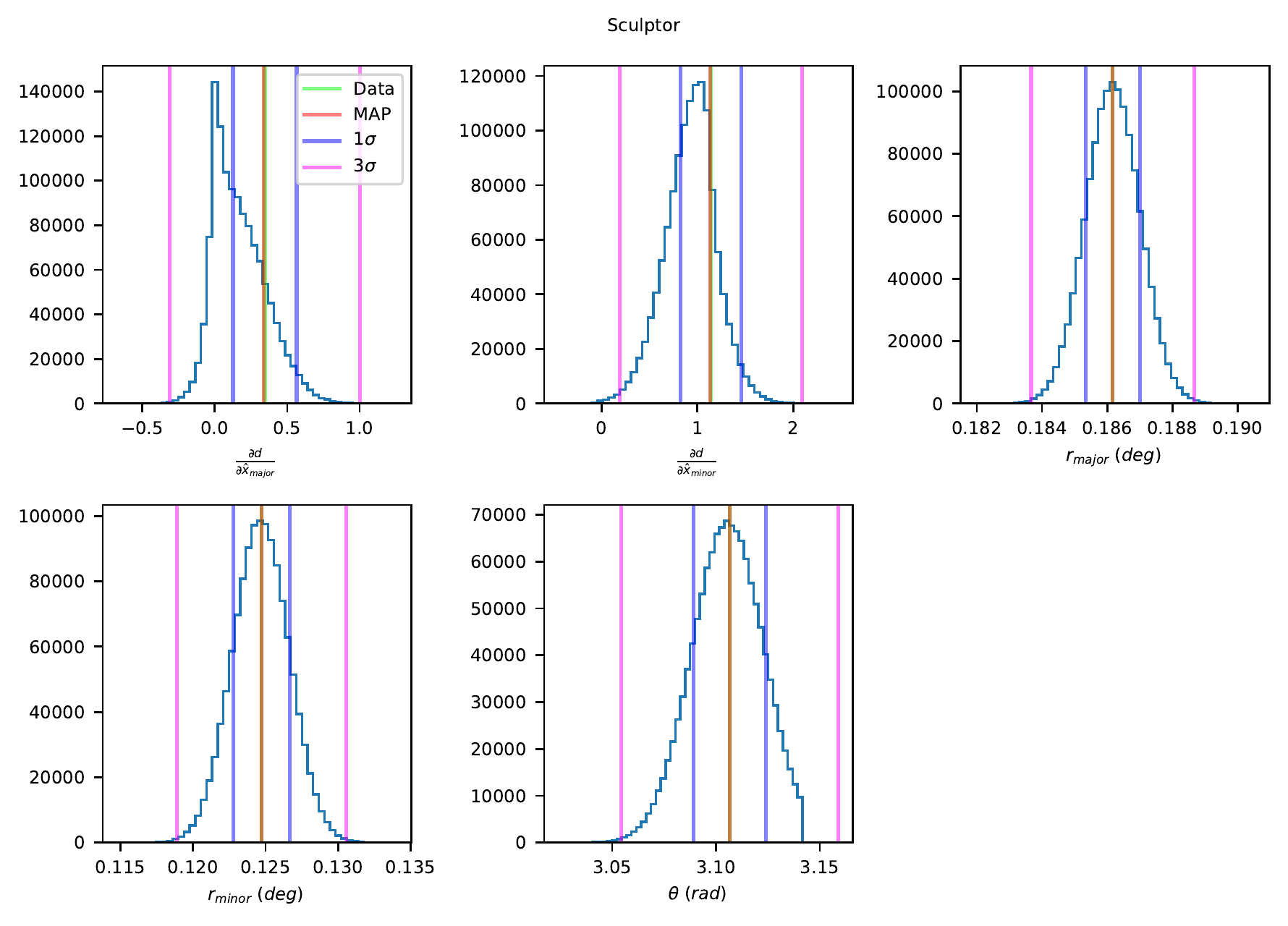}
    \caption{The plots are marginal histograms of parameters inferred from the posterior of Sculptor triaxial ellipsoid model, where $d$ is the distance in pc, $\hat x_{major}$ and $\hat x_{minor}$ is coordinate in semi-major/semi-minor coordinate system in pc, $r_{major}$ is half-light radius along semi-major axis, $r_{minor}$ is half-light radius along semi-minor axis for 2D Plummer and $\theta$ is position angle of major axis for 2D Plummer. The green lines show the values calculated from Sextans observed data, blue lines show their $1$\,$\sigma$ range and purple lines are $3$\,$\sigma$ range. The red line shows the parameters from the MAP for the triaxial ellipsoid model and is sitting on top of the green line.}
    \label{fig_sculptor_free_posterior_check}
\end{figure*}

Here we perform the posterior predictive check for the 3-D model fitting of Sextans and Sculptor. We sample from the posterior of 3-D model, then use analytic formula in Appendix~\ref{sec:3d_model_project} to calculate sample's $\hat r_{\text{major}}, \hat r_{\text{minor}}, \hat \theta_{\text{pos}}, \hat c_{x}$ and $\hat c_{y} $ which we defined in Section~\ref{sec:3d_model}, and compare the distribution of these parameters of samples with the same parameters of the data. Figure~\ref{fig_sculptor_free_posterior_check} shows the posterior predictive check results for the triaxial model of Sculptor.

We can see that the peak of distribution of gradient along semi-major is between $1\sigma$ and $2\sigma$ boundary and a small part of the distribution of gradient along semi-minor is outside $1\sigma$ boundary. The model is still acceptable because the distribution is still within $3\sigma$ boundary. Considering the triaxial model should be very flexible and should have the ability to perfectly fit the data, we also calculate a maximum a posteriori (MAP) model shown by the red line in the Figure~\ref{fig_sculptor_free_posterior_check} that also agrees well with data.

\begin{figure*}
    \includegraphics[width=\textwidth]{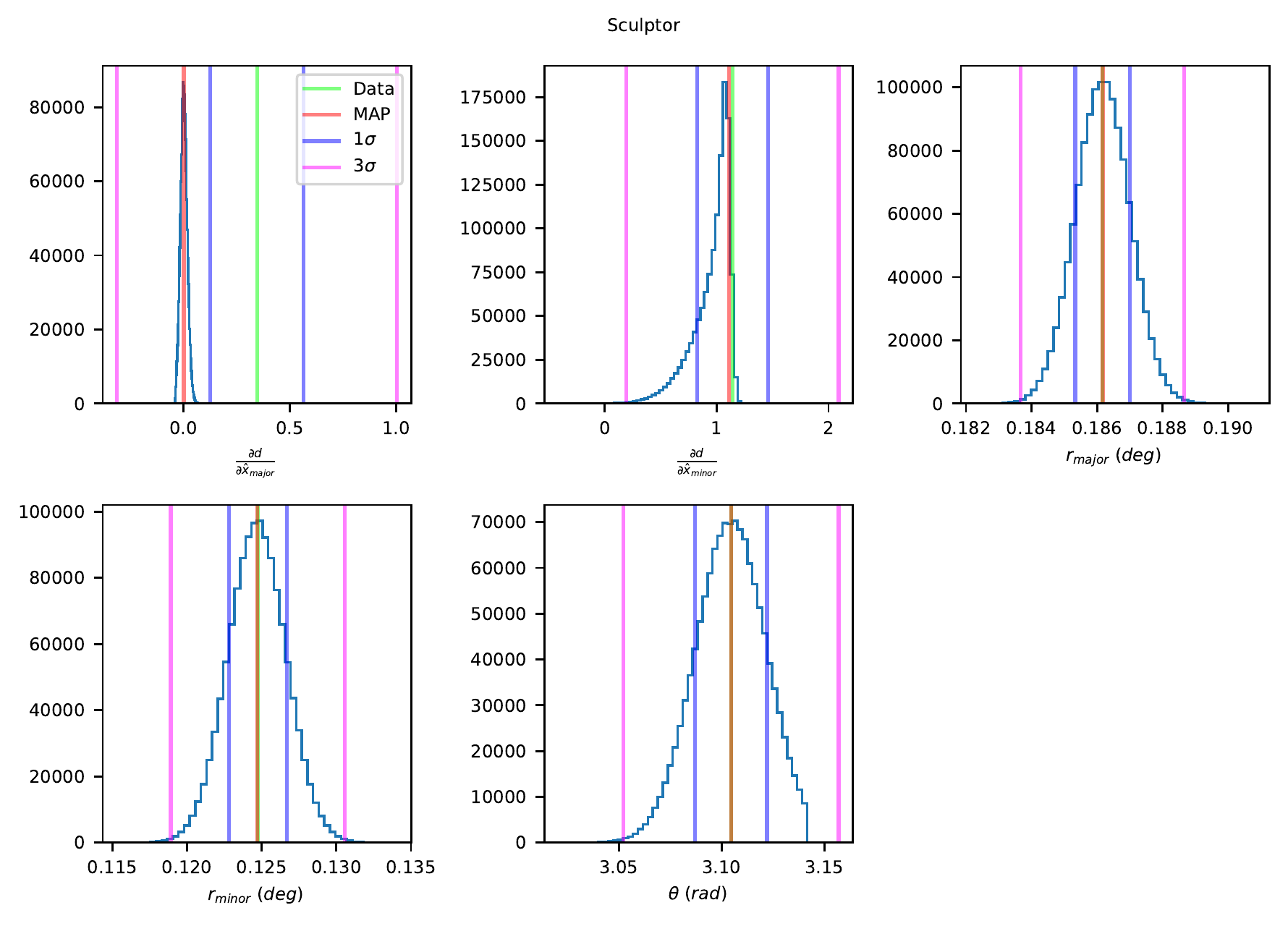}
    \caption{The plots are marginal histograms of parameters inferred from the posterior of Sculptor oblate ellipsoid model, where $d$ is the distance in pc, $\hat x_{major}$ and $\hat x_{minor}$ is coordinate in semi-major/semi-minor coordinate system in pc, $r_{major}$ is half-light radius along semi-major axis, $r_{minor}$ is half-light radius along semi-minor axis for 2D Plummer and $\theta$ is position angle of major axis for 2D Plummer. The green line is the value calculated from Sextans observed data, blue lines are $1\sigma$ boundary and purple lines are $3\sigma$ boundary. The red line is the result of MAP estimation for the oblate ellipsoid model.}
    \label{fig_sculptor_oblate_posterior_check}
\end{figure*}

For the oblate model, Figure~\ref{fig_sculptor_oblate_posterior_check} shows the result of posterior predictive check for Sculptor. The whole distribution of gradient along semi-major axis is outside $1\sigma$ boundary, and a small part of the distribution of gradient along semi-minor is outside  $1\sigma$ boundary. We are not surprised about the distribution of gradient along semi-major axis are tightly constrained because an oblate should have zero gradient along semi-major as discussed in Section~\ref{sec:3d_model}, and we can see that for oblate model, the MAP estimation cannot perfectly fit the data. This result is consistent with the distribution of triaxiality in the triaxial model fitting result, which shows zero triaxiality is possible but the triaxial model does not have a strong preference for oblate model.

Overall the result shows both the triaxial and the oblate models agree with the data of Sculptor. Based on posterior predictive check, we can accept the fitting results of both models.

We did the same analysis for Sextans and the posterior predictive check result is consistent with the data within $1\sigma$, we will not show the detail here.

\section{Extinction and Gradient}
\label{sec:extinction_check}

\begin{figure*}
    \includegraphics[width=8cm]{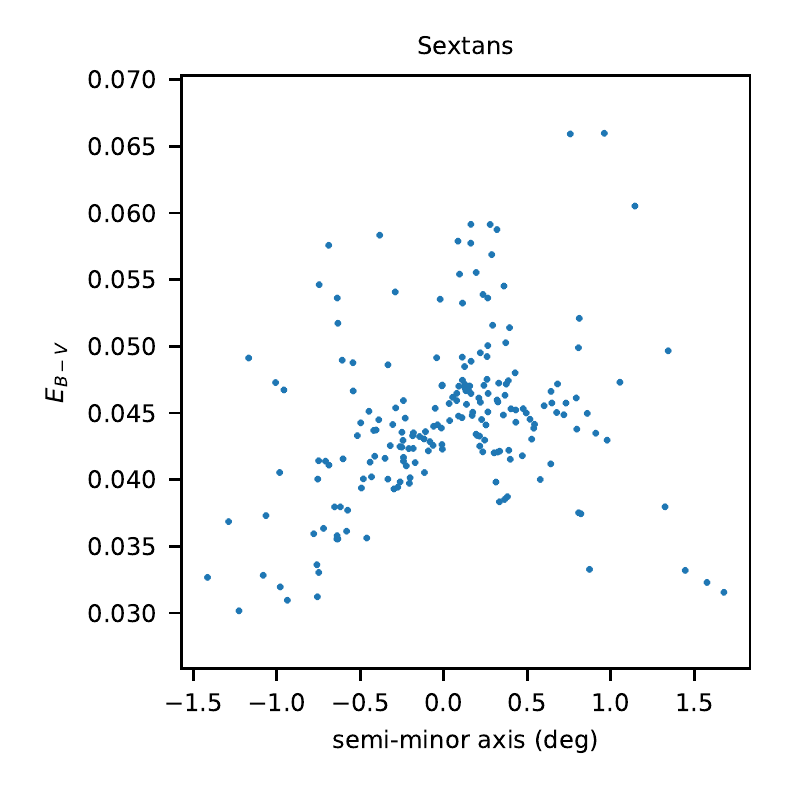}
    \includegraphics[width=8cm]{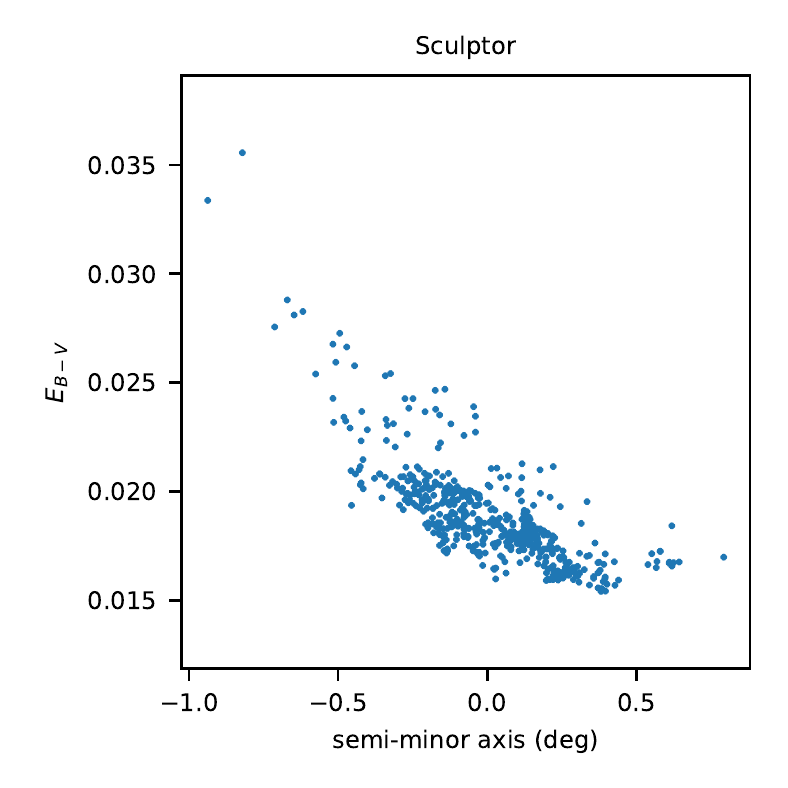}
    \caption{The plots shows the relationship between BHB stars' $E_{B-V}$ and their position along semi-minor axis in Sextans (left panel) and Sculptor (right panel), and the dots are BHBs we used in distance gradient fitting.}
    \label{fig_ebv_minor}
\end{figure*}

\begin{figure}
    \includegraphics[width=\columnwidth]{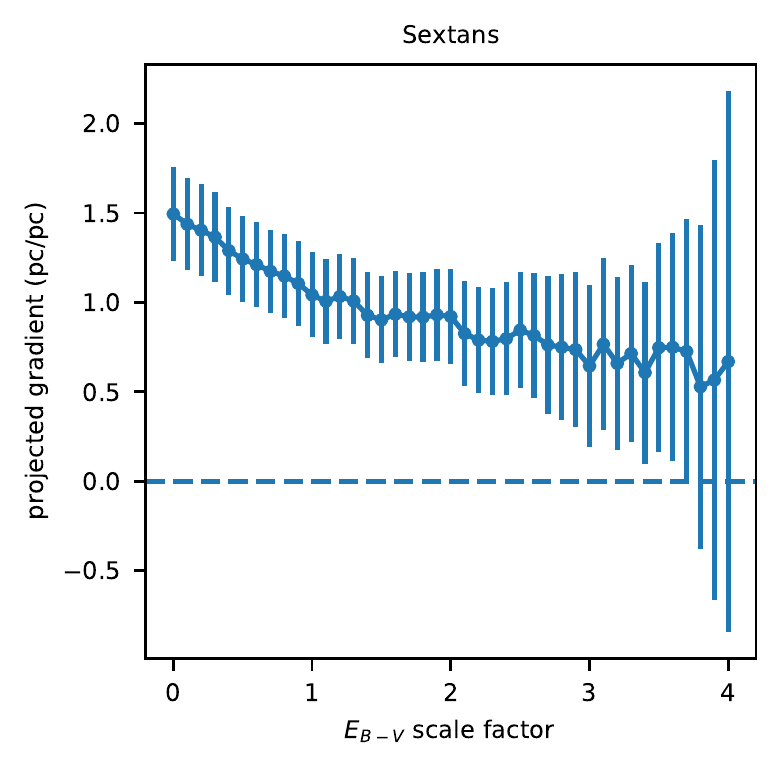}
    \caption{The plot shows the Sextans' projected distance gradient (pc/pc) along the direction from zero gradient to average gradient that we obtain with different $E_{B-V}$ scale factor. The errorbar is the standard deviation calculated after projection.
    }
    \label{fig_ebv_gradient_sextans}
\end{figure}

\begin{figure}
    \includegraphics[width=\columnwidth]{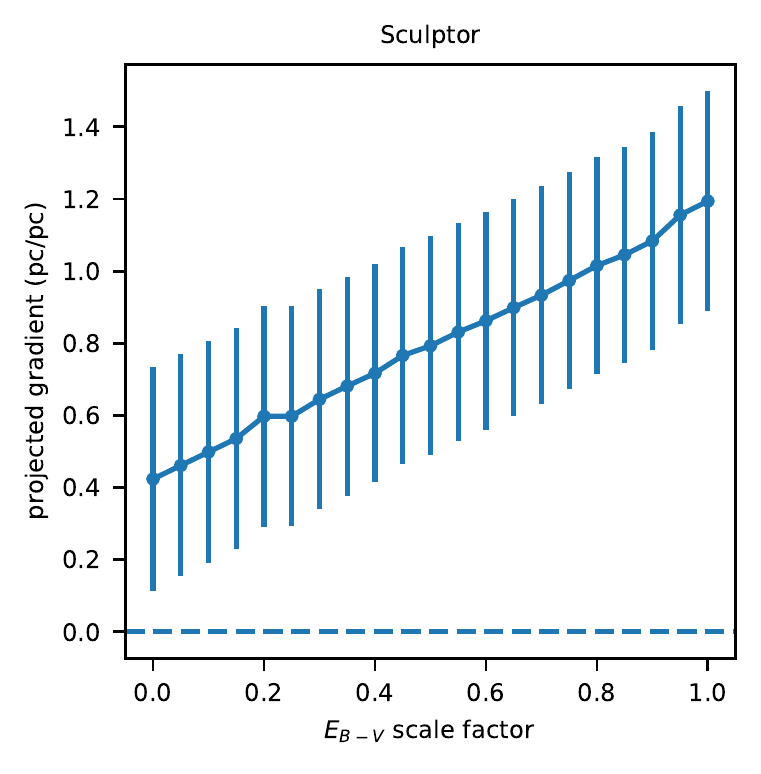}
    \caption{The plot shows the Sculptor' projected distance gradient (pc/pc) along the direction from zero gradient to average gradient that we obtain with different $E_{B-V}$ scale factor. The errorbar is the standard deviation calculated after projection.
    }
    \label{fig_ebv_gradient_sculptor}
\end{figure}

We notice that there is also $E_{B-V}$ gradient in the Sextans and Sculptor. Figure~\ref{fig_ebv_minor} shows the relationship between $E_{B-V}$ and position along semi-minor axis for BHB stars in Sextans (left panel) and Sculptor (right panel). It's clear to see the gradient. To study how the changes $E_{B-V}$ affect the distance gradient measurement, we scale $E_{B-V}$ by different factors. Figure~\ref{fig_ebv_gradient_sextans} shows the result for Sextans, where to eliminate the observed gradient  the extinction need to be scaled by almost 3.0. Figure~\ref{fig_ebv_gradient_sculptor} shows the result for Sculptor, it shows that we need to scale $E_{B-V}$ to zero to almost eliminate the distance gradient. These results show that it is extremely unlikely that the observed effects are caused by incorrect treatment of extinction. Even if there is a small change in extinction, that won't change the significance of our distance gradient measurement.

\bsp	
\label{lastpage}
\end{document}